\documentclass[12pt]{article}
\usepackage{amscd,amssymb,amsmath,latexsym,enumerate}
\usepackage[mathscr]{euscript}
\usepackage{graphicx}
\usepackage{mathrsfs}
\usepackage{epsfig}
\usepackage{fancybox}
\usepackage{verbatim}
\usepackage{tikz}
\usepackage{tikz-cd}
\usetikzlibrary{calc}
\usepackage{float}
\usepackage{todonotes}
\usepackage{multicol}
\usepackage{mathtools}
\usepackage[final]{pdfpages}

\usepackage{xcolor}
\definecolor{MyBlue}{cmyk}{1,0.13,0,0.63}
\definecolor{MyGreen}{cmyk}{0.91,0,0.88,0.52}
\newcommand{\mylinkcolor}{MyBlue}
\newcommand{\mycitecolor}{MyGreen}
\newcommand{\myurlcolor}{black}

\newcommand{\tick}[1]{\filldraw[black!100,line width=0.5mm,fill=none] {#1} ellipse (0.0 and 0.2)}

\usepackage{hyperref}
\hypersetup{%
  bookmarksnumbered=true,bookmarksopen=false,%
  plainpages=false,% necessary to prevent duplicate page identifiers
  linktocpage=true,%
  colorlinks=true,breaklinks=true,%
  linkcolor=\mylinkcolor,citecolor=\mycitecolor,urlcolor=\myurlcolor,%
  pdfpagelayout=OneColumn,%
  pageanchor=true,%
}

\usepackage{etoolbox}
\makeatletter
\patchcmd{\math@cr@@@align}{\cr}{\global\let\df@label\@empty\cr}{}{}
\makeatother

\title{Footprint of a topological phase transition \\ on the density of states}

\author{Joris De Moor$^1$, Christian Sadel$^2$, Hermann Schulz-Baldes$^1$
\\
\\
{\small $^1$Friedrich-Alexander-Universit\"at Erlangen-N\"urnberg}
\\
{\small Department Mathematik, Cauerstr.~11, D-91058 Erlangen, Germany}
\\
{\small $^2$Pontifica Universidad Cath\'olica de Chile}
\\
{\small Facultad de Matem\'aticas, Av. Vicu\~na Mackaenna 4860, Santiago 7820436, Chile}
}

\date{ }

\newtheorem{theorem}{Theorem}
\newtheorem{proposition}[theorem]{Proposition}
\newtheorem{lemma}[theorem]{Lemma}

\def\essinf{\mathop{\rm ess\,inf}}
\def\esssup{\mathop{\rm ess\,sup}}

\begin{document}

\maketitle

%%%%%%%%%%%%%%%%%%%%%%%%%%%%%%%%%%%%%%%%%%%%%%%%%%%%
\begin{abstract}
For a generalized Su-Schrieffer-Heeger model the energy zero is always critical and hyperbolic in the sense that all reduced transfer matrices commute and have their spectrum off the unit circle. Disorder driven topological phase transitions in this model are characterized by a vanishing Lyapunov exponent at the critical energy. It is shown that the integrated density of states away from a transition has a pseudogap with an explicitly computable H\"older exponent, while it has a characteristic divergence (Dyson spike) at the transition points. The proof is based on renewal theory for the Pr\"ufer phase dynamics and the optional stopping theorem for martingales of suitably constructed comparison processes.
\end{abstract}
%%%%%%%%%%%%%%%%%%

%\noindent {AMS MSC2020:} 82B44, 37H15, 39A21, 37H30
% 	82B44  	Disordered systems (random Ising models, random Schroedinger operators, etc.) in equilibrium statistical mechanics
%     37H15  Random dynamical systems aspects of multiplicative ergodic theory, Lyapunov exponents
%     39A21 Oscillation theory for difference equations
%     37H30   Stability theory for random and stochastic dynamical systems

%%%%%%%%%%%%%%%%%%%%%%%%%%%%%%%%%%
\section{Context and main result}

The SSH model (Su-Schrieffer-Heeger~\cite{SSH}) is the prototype of a chiral topological insulator in dimension one. Here a slightly generalized and disordered or {\it dirty} version of it will be considered. In such systems, one can associate a noncommutative winding number as a topolo\-gi\-cal invariant to the Fermi projection, provided that the Fermi level lies in a spectral region of Anderson localization. If one modifies the parameters of the system (such as the strength of the disorder in the hopping and on-site masses, see below), the topological invariant may change and the transition points make up the so-called topological phase boundary. It is known (Section~6.6 in~\cite{PS} and Section~5.5 in~\cite{ST}) that there is no dynamical Anderson localization for models on the phase boundary. In the disordered SSH model one can determine the phase boundary as those points at which the (smallest non-negative) Lyapunov exponent at energy $E_c=0$ vanishes~\cite{MSHP}. Away from these points, one can prove Anderson localization throughout the whole spectrum~\cite{Sh}. The novel contribution of this work is that the integrated density of states (IDOS) has a pseudo-gap at zero energy for parameters away from the phase boundary, while the density of states (DOS) has a characteristic divergence at the phase boundary. 

\vspace{.2cm} 

To formulate the main result, let us write out the generalized dirty SSH Hamiltonian $H$ to be considered here. Over each site of the lattice $\mathbb{Z}$, the system has a quantum cavity with $2L$ orbitals so that the total Hilbert space is $\ell^2(\mathbb{Z},\mathbb{C}^{2L})$. On each site acts a chiral symmetry operator $J=\mathrm{diag}(\mathbf{1}_L,-\mathbf{1}_L)$ which naturally extends to a symmetry on $\ell^2(\mathbb{Z},\mathbb{C}^{2L})$. Within the cavity over site $n$, the Hamiltonian is off-diagonal in the grading of $J$ with entry given by a random invertible matrix $M_{n}$. Furthermore, all sites are supposed to be connected by rank one operators $B$ with random couplings $t_n$. Hence the action of $H$ on $\psi=(\psi_{n})_{n\in\mathbb{Z}}\in\ell^2(\mathbb{Z},\mathbb{C}^{2L})$ is given by
\begin{equation}
\label{eq:SSH}
(H\psi)_{n}
\;=\;
-\,t_{n+1} 
\begin{pmatrix} 0 & B \\ 0 & 0 \end{pmatrix}
\psi_{n+1}
\,+\,
\begin{pmatrix} 0 & \!\!\! M_{n} \\ M_{n}^* & 0 \end{pmatrix}
\psi_{n}
\,-\,
\overline{t_{n}}
\begin{pmatrix} 0 & 0 \\ B^* & 0 \end{pmatrix}
\psi_{n-1}
\,.
\end{equation}
Here,  $t_{n}\in\mathbb{C}\setminus\{0\}$ with complex conjugate $\overline{t_n}$, and $t_n$, $M_{n}$ are random variables of the form $t_{n}=e^{\imath\phi_n}(1+\lambda\omega_n)$ and $M_{n}=\frac{1}{2}(m\,\mathbf{1}_L+\mu\omega'_n)$ where $\phi_n\in[0,2\pi)$,  $\omega_n\in [-\frac{1}{2},\frac{1}{2}]$, $\omega'_n\in \mathbb{C}^{L\times L}$. Then $\sigma_n=(\omega_n,\omega'_n)$ are random i.i.d. variables according to some compactly supported distribution.  For sake of concreteness, let us suppose that $B=e_1e_L^*$ where $\{e_1,\ldots,e_L\}$ is an orthonormal basis of $\mathbb{C}^L$. Moreover, $\lambda$, $\mu$ are coupling constants. Finally $m$ is a fixed mass parameter that assures that $M_{n}$ is invertible with a uniform lower bound on $M^*_nM_n$ (for $\mu$ sufficiently small) and it allows to drive the system into a topological phase. Note that $H$ also has the chiral symmetry $JHJ=-H$ so that the spectrum is symmetric around the center of band $E_c=0$. For reasons explained below, this energy will also be called critical. Just as any random Schr\"odinger operator, the generalized SSH model has a well-defined integrated density of states (IDOS) defined as the non-decreasing function $E\in\mathbb{R}\mapsto \mathcal{N}(E)$  given by
$$
\mathcal{N}(E)
\;:=\;
\lim_{N\to\infty}\frac{1}{N}\;\frac{1}{2L}\;\#\{\mbox{\rm eigenvalues of }H_N\,\leq\,E\}
\,,
$$
where $H_N$ is the restriction of $H$ to $\ell^2(\{1,\ldots,N\},\mathbb{C}^{2L})$.  The limit is known to exist almost surely~\cite{PF}. Furthermore, such a one-dimensional random model has a (smallest non-negative) Lyapunov exponent $\gamma(E)\geq 0$ for every energy $E\in\mathbb{R}$. Further down this will be introduced more carefully and it will also be shown that at the critical energy $\gamma(0)=|\mathbb{E}(\log(\kappa))|$ where $\kappa_{\sigma}$ is a positive random variable defined by
$$
\kappa_{\sigma}
\;=\;
\frac{1}{|e_1^* M_\sigma^{-1} e_L t_\sigma|}\,
\quad
\text{where}
\quad
t_\sigma=1+\lambda \omega
\,,\quad
M_\sigma=\frac12(m\mathbf{1}_L+\omega')
\,,
$$
with $\sigma=(\omega,\omega')\in \mathbb{R} \times \mathbb{C}^{L\times L}$ as in the model above. By definition (as in~\cite{MSHP}), the parameters $\lambda,\mu$ at which the Lyapunov exponent at the center of band vanishes make up  the phase boundary $\mathcal{P}$ of the SSH model, namely 
$$
\mathcal{P}
\;=\;
\big\{
(\lambda,\mu)\in\mathbb{R}^2\;:\;\gamma(0)=|\mathbb{E}(\log(\kappa))|=0
\big\}
\,.
$$
The main result of this work now states that one can read off the IDOS whether a model lies on $\mathcal{P}$ or not. 

%%%%%%%%%%%%%%%%%%%%%%%%%%%%%%%%%%%%%%%%%%%%%%%
\begin{theorem}
\label{theo:SSH}
For $(\lambda,\mu)\not\in\mathcal{P}$ lying off the phase boundary,  {\it i.e.} $\mathbb{E}(\log(\kappa))\neq 0$, the IDOS of the dirty SSH has a pseudo-gap at $0$ in the sense that 
\begin{equation}
\label{eq:pseudo}
\lim_{E \to 0}\, \frac{\log\big|\,\mathcal{N}(E)\,-\,\mathcal{N}(0)\,\big|}{\log(E)}
\;=\;
\nu
\,,
\end{equation}
where $\nu>0$ is determined as the unique positive solution of $\mathbb{E}(\kappa^{\nu})=1$ if $\mathbb{E} (\log(\kappa))<0$, and the solution of $\mathbb{E}(\kappa^{-\nu})=1$ otherwise. On the other hand, for $(\lambda,\mu)\in\mathcal{P}$ on the phase boundary, {\it i.e.} $\mathbb{E}(\log(\kappa))=0$, with some constant $C$, the DOS has a characteristic divergence at $0$
specified by
\begin{equation}
\label{ineq:log-div}
\left|\,\mathcal{N}(E)\,-\,\mathcal{N}(0)\,-\,\tfrac{1}{4L}\,\mathbb{E}\big(\big(\log(\kappa)\big)^2\big)\,\big(\log(E)\big)^{-2}\,\right|
\;\leq\;
C \;|\log(E)|^{-3}
\,.
\end{equation}
\end{theorem}
%%%%%%%%%%%%%%%%%%%%%%%%%%%%%%%%%%%%%%%%%%%%%%%

Let us compare Theorem~\ref{theo:SSH} with the literature on random hopping models which, as will be explained in Section~\ref{sec:transfer},  is essentially the particular case $L=1$ of the generalized SSH model. For the random hopping model, the upper bound $|\mathcal{N}(E)-\mathcal{N}(0)| \leq C_\delta |E|^{\nu-\delta}$ was proved in \cite{DS} for all $\delta>0$. Hence \eqref{eq:pseudo} provides also  the corresponding lower bound. For the random hopping model and points on $\mathcal{P}$, the characteristic divergence \eqref{ineq:log-div} is referred to as Dyson's spike, due to his work \cite{Dys} showing this for a particular distribution of the random hopping terms. Apart from Dyson's work, there are several non-rigorous works on both regimes covered by Theorem~\ref{theo:SSH}. Section~V.E in the review \cite{EM} contains relevant references. The behavior of the integrated  density of states as in \eqref{ineq:log-div} was more recently proved rigorously to hold under even weaker assumptions by Kotowski and Vir\'ag \cite{KV}  (no independence is assumed in their work, merely a sufficiently rapid correlation decay). However, no explicit error bound of order $|\log(E)|^{-3}$ was provided, merely a bound of order $o(E) \log(E)^{-2}$. Here we place all of these results in the joint context of topological phases and provide considerable technical improvements in the proofs. 

\vspace{.2cm}

In order to justify this last statement, let us provide a more detailed technical comparison with the work of Kotowski and Vir\'ag \cite{KV}. First of all, both works analyze the perturbation of the rotation number of the induced dynamics on projective space $\mathbb{P}(\mathbb{R}^2)\cong \mathbb{S}^1$ driven by the transfer matrices. The unperturbed dynamics at $E=0$ leaves two critical points and semicircles in between invariant.  The energy-dependent perturbation adds some rotation around the critical points (all in the same direction) and one needs to analyze the number of passages by the critical points into the next semi-circle. In the long-time limit one then obtains the rotation number which is equivalent to the density of states. In \cite{KV}  the free dynamics (which does not rotate) is subtracted by conjugations which involve products of the $\kappa$'s and these products can push the $\mathcal{O}(E)$ rotation induced by the perturbation.  Now the process is compared to some family of different dynamics (with an additional parameter $\delta$)  that are partially slower/faster (under certain conditions,  and for number of steps $n<\delta (\tfrac{1}{E})^{\delta/4}$ not too large). Then the number of crossings to the 'next' semi-circle is shown to be approximately equal to the number of times where the log-transformed free dynamics $\sum_{l=1}^n \log(\kappa_l)$ makes jumps of order $|\log(E)|$.  Playing with all the parameters one can get to some scaling limit for $E=e^{-\sqrt{n}} \to 0$ and $n \to \infty$.  Finally,  using probabilistic techniques ({\it cf.} \cite[Theorem 3.10]{KV})  the authors obtain bounds for the rotation number for $E$ small, meaning they can let $E$ fixed and $n \to \infty$, namely a statement similar to \eqref{ineq:log-div}, but only with an error $o(E)$. 

\vspace{.1cm}
 
In contradistinction, in this work the dynamics is analyzed directly, without conjugation of the free dynamics.  It is then compared to suitable slower and faster dynamics which allow to estimate the crossings at fixed $E$.  In essence, the slower dynamics drops the $\mathcal{O}(E)$ rotation except for the region close to the critical points, and the faster dynamics essentially replaces the $\mathcal{O}(E)$ perturbations by a $o(E)$ drift going forward. Formulating the rotation number in terms of an expectation of a certain stopping time, one can use the optional stopping theorem to get the claimed estimates. The present approach is more direct and a lot less technical than the one of \cite{KV}. Moreover, the constructions work immediately for both cases treated in Theorem~\ref{theo:SSH}; only the constructed martingales and the consequent usage of the optional stopping theorem are of different nature, leading to the different behavior at $E=0$. It is hard to see how to modify the techniques of $\cite{KV}$ for the case $\mathbb{E}(\log(\kappa))\neq 0$ (which does not mean that it cannot be done).

\vspace{.2cm}

To conclude this introductory section, let us mention that we are currently investigating several interesting open questions on the generalized SSH model. First of all, one would like to have a controlled perturbation theory for the Lyapunov exponent in the vicinity of the critical energy of models on the transition (as in \cite{PF,JSS,DKS}). This pends on a good understanding of the Furstenberg measure. As illustrated numerically in Figure~\ref{fig:IDOS-Lyap} below, the Lyapunov exponent (or inverse localization length) has a similar singular behavior as the IDOS, as predicted by theoretical physicists (see \cite{EM}). Second of all, we expect that all these states at energies with large localization length (as exhibited in Theorem~\ref{theo:SSH}) lead to a quantitative lower bound on the quantum dynamics (going beyond the statements of~\cite{PS,ST} showing that models at the topological phase boundaries cannot be dynamically Anderson localized). The mechanism behind this quantitative delocalization phenomenon is similar as in the random dimer model~\cite{DWP}, random polymer model~\cite{JSS} or the random Kronig-Penney model~\cite{DKS}, but a proof is much more subtle due to the presence of the singularities of the DOS and the Lyapunov exponent. Finally, another question concerns the fate of the (likely enhanced) area law in these models \cite{MPS}. Let us not that the nature of the level statistics near the critical energy for models on the transition was already determined in \cite{KV}.

%%%%%%%%%%%%%%%%%%%%%%%%%%%%%%%%%%
\section{Transfer matrices and critical energies}
\label{sec:transfer}

The proof of Theorem~\ref{theo:SSH} uses the transfer matrix formalism for the study of quasi-one-dimensional Jacobi operators. Clearly, the SSH Hamiltonian \eqref{eq:SSH} is a such a block Jacobi matrix with $2L\times 2L$ block entries on every site. However, the off-diagonal entries are not invertible so that one cannot define the $4L\times 4L$ transfer matrices in the usual form (which involves working with the inverse of the off-diagonal terms). One rather has to pass to the so-called reduced transfer matrices~\cite{DC,Sad,SB}. In the present situation, the matrices $B$ are of rank $1$ and therefore the reduced transfer matrices will be of size $2\times 2$ satisfying 
\begin{equation}
\label{eq:symplectic}
T^*\,I\,T
\;=\;
I
\,,
\qquad
I
\;:=\;
\begin{pmatrix}
0 & -1 \\ 1 & 0
\end{pmatrix}
\,.
\end{equation}
For their construction, let us denote by  $\lbrace e_1,\ldots,e_{L} \rbrace$ a basis for $\mathbb{C}^{L}$, chosen such that $B=e_1e_L^*$. Then the ranges of the lower and upper entries in the block Jacobi matrices both have a one-dimensional span $\mathcal{H}^+=\mathrm{span}\{e_1\}$ and $\mathcal{H}^-=\mathrm{span}\{e_L\}$ in $\mathbb{C}^L$, respectively. These two spaces are orthogonal as required in~\cite{SB}. The relevant part of the resolvent of the diagonal part is
$$
\binom{e_1}{e_L}^*
\left(E\,\mathbf{1}\,-\,
\begin{pmatrix} 0 & \!\! M_{n} \\ M_{n}^* & 0 \end{pmatrix}
\right)^{-1}
\binom{e_1}{e_L}
\;=\;
\begin{pmatrix}
G^{E,-,-}_n & G^{E,-,+}_n \\ G^{E,+,-}_n & G^{E,+,+}_n 
\end{pmatrix}
\,,
$$
by definition of the $4$ scalar entries on the r.h.s.. Therefore the reduced transfer matrices, given in (15) or (17) of~\cite{SB},  are 
\begin{equation}
\label{eq:factorized}
T^E_n
\;=\;
-\,
\begin{pmatrix}
(G^{E,-,+}_n)^{-1} & (G^{E,-,+}_n)^{-1}G^{E,-,-}_n\\
-G^{E,+,+}_n(G^{E,-,+}_n)^{-1}& G^{E,+,-}_n-G^{E,+,+}_n ( G^{E,-,+}_n)^{-1}G^{E,-,-}_n 
\end{pmatrix}
\begin{pmatrix}
\frac{1}{t_{n}} & 0\\
0 & \overline{t_{n}}
\end{pmatrix}
\,.
\end{equation}
Let us note that this SSH model also fits into the scheme of one-channel operators and the transfer matrices above correspond exactly to (1.10) in \cite{Sad}. The invertibility of $M_n$ assures that there is an open interval around $E_c=0$ in which $\binom{0 \;\;\;M_n}{M_n^*\;\;\;0}$ has no eigenvalue and therefore all three hypothesis stated in~\cite{SB} are satisfied and one can conclude that the reduced transfer matrices are analytic in $E$ in this interval. As already stressed above, they also satisfy~\eqref{eq:symplectic}. This implies that their determinant is of unit modulus. In order to attain a determinant equal to $1$, one can use an energy dependent gauge transformation $(W^E\psi)_n= e^{\imath\varphi_n}\psi_n$ with phases $\varphi_n\in[0,2\pi)$ to be chosen next.  In fact, the Hamiltonian $W^E H (W^E)^*$ is of the same block triagonal form as $H$;  the diagonal entries $M_n$ are unchanged, but the the off-diagonal entries are obtained by replacing $t_n$ by $t_ne^{\imath\,\delta\varphi_n}$ with $\delta\varphi_n=\varphi_n-\varphi_{n-1}$. According to~\eqref{eq:factorized} this changes $\det(T^E_n)$ to $e^{-2\imath\,\delta\varphi_n}\det(T^E_n)$. Therefore for every energy $E\in\mathbb{R}$ the angles $\varphi_n$ can be chosen iteratively in $n$ such that all transfer matrices at $E$ have unit determinant. Hence in the following, we will always assume $\det(T^E_n)=1$ without further mention of the gauge transformation (the necessary phases are simply absorbed in redefined $t_n$). This implies that the whole reduced transfer matrix $T^E_n$ is real, as one readily deduces from the relation $T^*=IT^{-1}I^*$ following from~\eqref{eq:symplectic}. Alternatively, one can simply assume that the $t_n$ and matrices $M_n$ are all real, then also the reduced transfer matrices in~\eqref{eq:factorized} are real with unit determinant (or equal to $-1$, which can again be absorbed by a gauge transformation).  When focusing on the behavior of the reduced transfer matrices at $E_c=0$, one needs the relations 
$$
G^{0,-,-}_n
\;=\;
0\,,
\qquad
G^{0,+,-}_n
\;=\;
-e_1^*(M_{n})^{-1}e_L\,,
\qquad 
G^{0,-,+}_n 
\;=\;
(G^{0,+,-}_n)^*\,,
\qquad
G^{E,+,+}_n 
\;=\;
0
\,.
$$
They imply
\begin{equation}
\label{eq:T^0}
T^0_n
\;=\;
-\,
\begin{pmatrix}
\kappa_n & 0\\
0 & \frac{1}{\kappa_n}
\end{pmatrix}
\,,
\qquad
\kappa_n
\;=\;
\frac{1}{G^{0,-,+}_n\,t_{n}}
\,.
\end{equation}
Hence $E_c=0$ is indeed a hyperbolic critical energy in the sense~\cite{DS} that all transfer matrices commute and some of them are hyperbolic (if any of the distributions is non-trivial so that $|\kappa_n|$ is not identically equal to $1$). Based on this, one can further expand the reduced transfer in $E$ which leads to the situation \eqref{eq:expansion} studied in the remainder of the paper.

\vspace{.2cm}

Let us briefly specify how to obtain the model studied in~\cite{MSHP} as well as the random hopping model from \cite{Dys,KV,DS}. One chooses $L=1$, $B=1$ and then the matrices $M_n$ are scalars $m_n$. If one denotes $\lambda=W_1$, $\mu=W_2$, and $\omega_n$ and $\omega'_n$ have a uniform distribution on $[-\frac{1}{2},\frac{1}{2}]$, one obtains after conjugation with a suitable Cayley transform exactly the random Hamiltonian of~\cite{MSHP}. For these particular distributions, the Lyapunov exponent at $E_c=0$ can be calculated explicitly~\cite{MSHP}, but the results of this paper do not depend on these particular choices. Let us also spell out the reduced transfer matrices in this case. One finds that $G^{E,+,-}_n =\frac{m_n}{E^2-m_{n}^2}$ and $G^{E,+,+}_n =G^{E,-,-}_n =\frac{E}{E^2-m_{n}^2}$. Then one readily checks 
$$
T^E_n
\;=\;
\begin{pmatrix}
\frac{m_{n}}{t_{n}} & 0\\
0 & \frac{t_{n}}{m_{n}}
\end{pmatrix}
\,+\,E
\begin{pmatrix}
0 & -\,\frac{t_{n}}{m_{n}}\\
\,\frac{1}{m_{n}\,t_{n}} & 0
\end{pmatrix}
\,-\,E^2
\begin{pmatrix}
\frac{1}{m_{n}\,t_{n}}& 0\\
0 & 0
\end{pmatrix}
\,.
$$
This is actually also connected to Dyson's random hopping model studied in \cite{Dys,KV}. More precisely, set $\hat{t}_{2n}=t_n$ and $\hat{t}_{2n+1}=m_n$ and suppose that they are identically distributed, then one can check
$$
T^E_n
\;=\;
\begin{pmatrix}
-\,E\,\tfrac{1}{\hat{t}_{2n+1}} & -\hat{t}_{2n+1} \\ \tfrac{1}{\hat{t}_{2n+1}} & 0
\end{pmatrix}
\begin{pmatrix}
-\,E\,\tfrac{1}{\hat{t}_{2n}} & -\hat{t}_{2n} \\ \tfrac{1}{\hat{t}_{2n}} & 0
\end{pmatrix}
\,,
$$
which is indeed the two-step transfer matrix of a random hopping Hamiltonian on $\ell^2(\mathbb{Z})$ given by
$$
(H\psi)_n
\;=\;
-\,\hat{t}_{n+1} \psi_{n+1}\,-\,\hat{t}_{n}\psi_{n-1}
\,,
\qquad
\psi=(\psi_n)_{n\in\mathbb{Z}}\in\ell^2(\mathbb{Z})
\,.
$$
Hence both the dirty SSH Hamiltonian from \cite{MSHP} and the random hopping model are particular cases of the generalized SSH Hamiltonian \eqref{eq:SSH}.

%%%%%%%%%%%%%%%%%%%%%%%%%%%%%%%%%%
\section{Pr\"ufer phase formalism near critical energy}

In the theory of products of random $2\times 2$ matrices, the associated Lyapunov exponent can be accessed via the random action of the  matrices on projective space, which in turn is bijectively mapped to a unit circle making up the so-called Pr\"ufer phases. By a Cayley transform, they are mapped to real numbers which are then called Dyson-Schmidt variables. In both cases, the action is implemented by a M\"obius transformation. This way of approaching the Lyapunov exponent is particularly efficient for perturbative expansions~\cite{PF,Luc}.  Furthermore, if the random matrices are the transfer matrices from a given one-dimensional random operator and the Pr\"ufer phases are suitably lifted to $\mathbb{R}$, then oscillation theory also allows to extract the DOS from the Pr\"ufer phases and again this is a good way to tackle perturbative problems. 

\vspace{.2cm}

Traditionally, perturbation theory is done in a coupling constant of the randomness, corresponding to a weak coupling limit of the randomness ({\it e.g.} in the one-dimensional Anderson model). However, there are other situations where the perturbative parameter is the energy distance to some critical energy, and is hence intrinsic to the model rather than an external parameter. The first example of this type is the random dimer model~\cite{DWP} and its generalization, the random polymer model~\cite{JSS}. In these models exists a so-called critical energy at which all (random) transfer matrices commute and, moreover, the spectrum of all these matrices lies on the unit circle so that they can simultaneously diagonalized into (random) rotations. Due to the latter fact, the critical energy of this type is called elliptic. On the other hand, in a random Kronig-Penney model there can be a critical energy at which the transfer matrices are all similar to a Jordan block~\cite{DKS}, so that the critical energy is then called parabolic. Finally, it was pointed out in~\cite{DS} that the random hopping model and the SSH model have a hyperbolic critical energy with transfer matrices having their spectra off the unit circle. Sections~\ref{sec:lower} and \ref{sec:upper} treat the case in which the Lyapunov exponent is nonvanishing at the critical energy, which is then called unbalanced. Section~\ref{sec:balanced} then concerns the so-called balanced case with a vanishing Lyapunov exponent.

\vspace{.2cm}

In order to cover other possible applications of hyperbolic critical energies and to stress structural aspects, let us consider the same set-up as in~\cite{DS}. Suppose $(\Sigma,{\bf p})$ is a compact probability space and $\sigma\in\Sigma\mapsto T^{E_c+\epsilon}_\sigma\in\mbox{\rm SL}(2,\mathbb{R})$ a family of transfer matrices over polymer blocks of length $L_\sigma\in\mathbb{N}$ which is of the form
\begin{equation}
\label{eq:expansion}
T^{E_c+\epsilon}_{\sigma}
\;=\;
\pm
\left[\mathbf{1}\,+\,a_\sigma\epsilon
\begin{pmatrix}
0 & -1 \\
1 & 0
\end{pmatrix}
\,+\,b_\sigma\epsilon
\begin{pmatrix}
0 & 1 \\
1 & 0
\end{pmatrix}
\,+\,c_\sigma\epsilon
\begin{pmatrix}
1 & 0 \\
0 & -1
\end{pmatrix}
\,+\,\mathcal{O}(\epsilon^2)\right]
D_{\kappa_{\sigma}}
\,.
\end{equation}
Here $a_\sigma,b_\sigma,c_\sigma$ are real numbers,  $\kappa_{\sigma}>0$ and furthermore
$$
D_{\kappa}
\;=\;
\begin{pmatrix}
\kappa & 0 \\ 0 &\frac{1}{\kappa}
\end{pmatrix}
\,.
$$
The hyperbolic critical energy will be called unbalanced if $\mathbb{E}(\log(\kappa))\not=0$ and balanced if $\mathbb{E}(\log(\kappa))=0$. In the former situation, we will always focus on the case $\mathbb{E}(\log(\kappa))<0$, as otherwise one can simply conjugate by the matrix $\binom{0\;-1}{1\;\phantom{-}0}$. The particular form \eqref{eq:expansion} covers the reduced transfer matrices $T^E_n$ of the generalized SSH model given in \eqref{eq:factorized} due to \eqref{eq:T^0} and the analyticity at $E_c=0$. It is also readily possible to deduce the random real coefficients for these models. In more general situations (such as random polymer models) one may use so-called modified transfer matrices to attain~\eqref{eq:expansion}, see~\cite{JSS,DS} for details. Let us note that one can show (see Proposition~3 in~\cite{DS}) that the inequalities $a_{\sigma}\geq 0$ and $a^2_\sigma\geq b_\sigma^2+c_\sigma^2$ hold for all $\sigma\in\Sigma$. In all arguments below, it is possible to absorb the contribution of $c_\sigma$ in the diagonal term by replacing $\kappa_{\sigma}$ by $\kappa_{\sigma}(1+\epsilon c_\sigma)$. In order to somewhat simplify notations, we will suppose $c_\sigma=0$ for all $\sigma\in\Sigma$. It will be useful to rewrite~\eqref{eq:expansion} as 
\begin{equation}
\label{eq:factorization}
T^{E_c+\epsilon}_{\sigma}
\;=\;
R^\epsilon_{\sigma}\,D_{\kappa_{\sigma}}
\,,
\end{equation}
with the notations 
$$
R^{\epsilon}_{\sigma}
\;=\;
\mathbf{1}\,+\,a_\sigma\epsilon
\begin{pmatrix}
0 & -1 \\
1 & 0
\end{pmatrix}
\,+\,b_\sigma\epsilon
\begin{pmatrix}
0 & 1 \\
1 & 0
\end{pmatrix}
\,+\,
\epsilon^2 \,A^\epsilon_\sigma\,,
\qquad
A_{\sigma}^{\epsilon}
\;=\;
\begin{pmatrix}
\alpha_{\sigma}^{\epsilon} & \beta_{\sigma}^{\epsilon} \\ \gamma_{\sigma}^{\epsilon} & \delta_{\sigma}^{\epsilon}
\end{pmatrix}
\,.
$$
The overall sign in~\eqref{eq:expansion} is neglected as it merely leads to a shift by $\pi$ in the Pr\"ufer phase dynamics below that is irrelevant for the  Pr\"ufer phases relative to the critical energy. 

\vspace{.2cm}

\begin{figure}
\centering
\includegraphics[width=8.4cm]{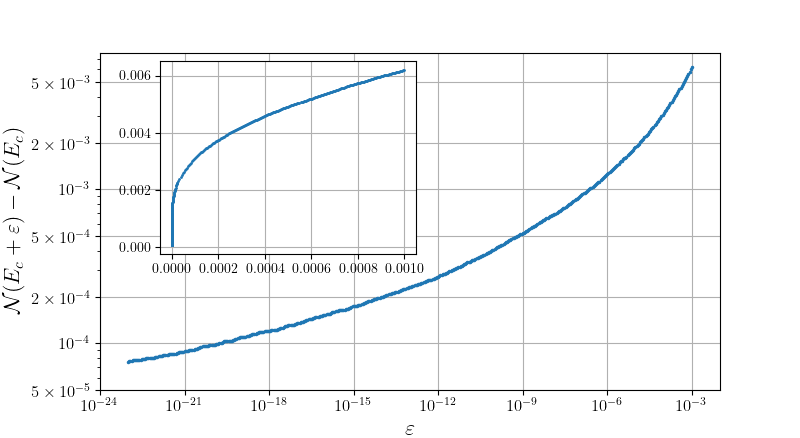}
\hspace{-.1cm}
\includegraphics[width=8.4cm]{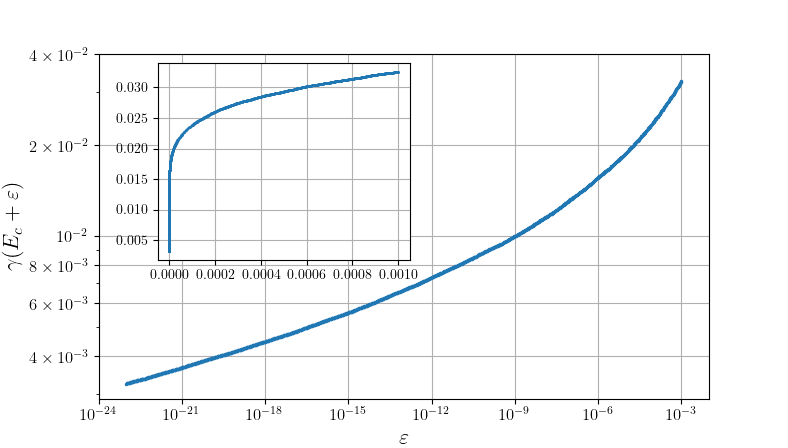}
\caption{\sl Numerical plot of the IDOS $\mathcal{N}(\epsilon)-\mathcal{N}(0)=\mathcal{N}(\epsilon)-\frac{1}{2}$ relative to the center of band $E_c=0$ and the Lyapunov exponent $\gamma(\epsilon)$ for the balanced random hopping model, both in a log-log plot. All points on these curves are computed via the Birkhoff sums~\eqref{eq:IDOS-Birkhoff} and~\eqref{eq:Lyap-Birkhoff} over orbits of length $N=10^7$.}
\label{fig:IDOS-Lyap}
\end{figure}

In the following, let us consider a random polymer Hamiltonian with hyperbolic critical energy so that the $n$th (possibly modified) transfer matrices are of the form~\eqref{eq:factorization} with coefficients drawn from the probability space $(\Sigma,\mathbf{p})$ (in which the $\kappa$ coefficients are taken to be independently and identically distributed). Hence $\omega=(\sigma_n)_{n\in\mathbb{Z}}$ is a configuration from $\Omega=\Sigma^\mathbb{Z}$.  The expectations w.r.t. the probability measure $\mathbb{P}$ on $\Omega$ will be denoted by $\mathbb{E}$. Associated are random coefficients and matrices $a_{\sigma_n}$, $b_{\sigma_n}$, $T^\epsilon_{\sigma_n}$, $\kappa_{\sigma_n}$, {\it etc.}, which for sake of notational convenience will simply be denoted by $a_n$, $b_n$, $T^\epsilon_n$, $\kappa_n$, {\it etc.}, unless there is some danger of misunderstanding. Associated to each configuration is a random sequence of Pr\"ufer phases $\theta^{\epsilon}_{n}\in\mathbb{R}$ at $\epsilon$ (and relative to the critical energy $E_c$) which can be introduced by
$$
e_{\theta^\epsilon_n}
\;=\;
\frac{T^{\epsilon}_n\,
e_{\theta^\epsilon_{n-1}}}{
\|T^{\epsilon}_n\,
e_{\theta^\epsilon_{n-1}}\|}
\,,
\qquad
e_{\theta}
\;:=\;
\begin{pmatrix} \cos (\theta)
\\ \sin(\theta) 
\end{pmatrix}
\,,
$$
\noindent a given (and irrelevant) initial condition $\theta^\epsilon_0$ and the lifting condition $\theta^{\epsilon}_{n+1} - \theta^{\epsilon}_{n}\in (-\frac{\pi}{2},\frac{3\pi}{2})$ fixing the branch. Note that this definition is induced by a group action of $\mbox{SL}(2,\mathbb{R})$ on $\mathbb{R}$ and hence $(\theta^{\epsilon}_{n})_{n\in\mathbb{Z}}$ is a Markov process on $\mathbb{R}$. As explained in detail in~\cite{JSS} and~\cite{DS}, the IDOS of the random polymer model is then given by
$$
\mathcal{N}(E_c+\epsilon) 
\;=\; 
\mathcal{N}(E_c)
\,+\,
\frac{1}{\pi}\,\frac{1}{\mathbb{E}(L_\sigma)}\;\lim_{N \to \infty}\,\frac{1}{N}\,\mathbb{E}(\theta^\epsilon_N)
\,.
$$
For the generalized SSH model this also holds by combining the arguments of~\cite{DS} with the oscillation theory as described in~\cite{SB}. The r.h.s. is the so-called rotation number, here relative to the critical energy. It is helpful to write it as a Birkhoff sum
\begin{equation}
\label{eq:IDOS-Birkhoff}
\mathcal{N}(E_c+\epsilon) 
\,-\, 
\mathcal{N}(E_c)
\;=\;\frac{1}{\pi}\,\frac{1}{\mathbb{E}(L_\sigma)}\;
\lim_{N \to \infty} \,\frac{1}{N}\,
\sum_{n=1}^N
\mathbb{E}(\theta^\epsilon_n\,-\,\theta^\epsilon_{n-1})
\,,
\end{equation}
because by the above each summand then lies in the interval $(-\frac{\pi}{2},\frac{3\pi}{2})$ and is called a phase shift. Before going into an intuitive description of the random dynamics of Pr\"ufer phases, let us furthermore recall from~\cite{JSS,DS} that the Lyapunov exponent can be expressed as a Birkhoff sum of the Pr\"ufer phases as well:
\begin{equation}
\label{eq:Lyap-Birkhoff}
\gamma(E_c+\epsilon) 
\;=\;
\lim_{N \to \infty} \,\frac{1}{N}\,
\sum_{n=1}^N
\mathbb{E}\big(\log(\|T^\epsilon_{n+1}e_{\theta^\epsilon_n}\|)\big)
\,.
\end{equation}
(One may include a factor $\frac{1}{\mathbb{E}(L_\sigma)}$ here.) The two formulas \eqref{eq:IDOS-Birkhoff} and \eqref{eq:Lyap-Birkhoff} allow to numerically compute the IDOS and the Lyapunov exponent for the random hoping model with great precision. As an example, both formulas are implemented in the balanced case of the random hopping model in Figure~\ref{fig:IDOS-Lyap}. In particular, this illustrates \eqref{ineq:log-div} and shows that the Lyapunov exponent has a similar behavior, as argued in the physics literature (see again Section V.E in \cite{EM}).

\vspace{.2cm}

\begin{figure}
	\begin{center}
		\begin{tikzpicture}[line join = round, line cap = round]
		\coordinate (a) at (0.0,0.8);
		\coordinate (b) at (0.0,-0.4);
		\coordinate (br) at (0.3,-0.3);
		\coordinate (bl) at (-0.3,-0.3);
		\coordinate (l) at (-1,0);
		\coordinate (m0) at (0,0);
		\tick{(m0)};
		\coordinate[label=above:{$\epsilon$}, label=below:{$\curvearrowright$}] (m0a) at ($(m0) + (a)$);
		\coordinate[label=below:{$0$}] (m0b) at ($(m0) + (b)$);
		\coordinate (m0br) at ($(m0) + (br)$);
		\coordinate (m1) at (1.5,0);
		\tick{(m1)};
		\coordinate[label=above:{$\epsilon$}, label=below:{$\curvearrowright$}] (m1a) at ($(m1) + (a)$);
		\coordinate[label=below:{$\frac{\pi}{2}$}] (m1b) at ($(m1) + (b)$);
		\coordinate (m1bl) at ($(m1) + (bl)$);
		\coordinate (m1br) at ($(m1) + (br)$);
		\coordinate (m2) at (3,0);
		\tick{(m2)};
		\coordinate[label=above:{$\epsilon$}, label=below:{$\curvearrowright$}] (m2a) at ($(m2) + (a)$);
		\coordinate[label=below:{$\pi$}] (m2b) at ($(m2) + (b)$);
		\coordinate (m2bl) at ($(m2) + (bl)$);
		\coordinate (m2br) at ($(m2) + (br)$);
		\coordinate (m3) at (4.5,0);
		\tick{(m3)};
		\coordinate[label=above:{$\epsilon$}, label=below:{$\curvearrowright$}] (m3a) at ($(m3) + (a)$);
		\coordinate[label=below:{$\frac{3\pi}{2}$}] (m3b) at ($(m3) + (b)$);
		\coordinate (m3bl) at ($(m3) + (bl)$);
		\coordinate (m3br) at ($(m3) + (br)$);
		\coordinate (m4) at (6.0,0);
		\tick{(m4)};
		\coordinate[label=above:{$\epsilon$}, label=below:{$\curvearrowright$}] (m4a) at ($(m4) + (a)$);
		\coordinate[label=below:{$2\pi$}] (m4b) at ($(m4) + (b)$);
		\coordinate (m4bl) at ($(m4) + (bl)$);
		\coordinate (m4br) at ($(m4) + (br)$);
		\coordinate (m5) at (7.5,0);
		\tick{(m5)};
		\coordinate[label=above:{$\epsilon$}, label=below:{$\curvearrowright$}] (m5a) at ($(m5) + (a)$);
		\coordinate[label=below:{$\frac{5\pi}{2}$}] (m5b) at ($(m5) + (b)$);
		\coordinate (m5bl) at ($(m5) + (bl)$);
		\coordinate[label=right:{$\theta$}] (r) at (8.5,0);
		\draw [->,color=black,line width=0.3mm] (l)--(r);
		\draw [->,color=black,line width=0.3mm] (m1bl)--(m0br);
		\draw [->,color=black,line width=0.3mm] (m1br)--(m2bl);
		\draw [->,color=black,line width=0.3mm] (m3bl)--(m2br);
		\draw [->,color=black,line width=0.3mm] (m3br)--(m4bl);
		\draw [->,color=black,line width=0.3mm] (m5bl)--(m4br);
		\end{tikzpicture}
		\caption{\it The dynamics $\theta^\epsilon_n$ for $\epsilon=0$ has fixed points at $0\,\mbox{\rm mod}\,\frac{\pi}{2}$ and therefore has many invariant intervals. However, for $\epsilon>0$ the dynamics crosses the fixed points by steps of size $\epsilon$ to the right. The hypothesis $\mathbb{E}(\log(\kappa_0)) < 0$ induces local drifts indicated by the arrows below the axis.}
		\label{fig:theta-unbalanced}
	\end{center}
\end{figure}
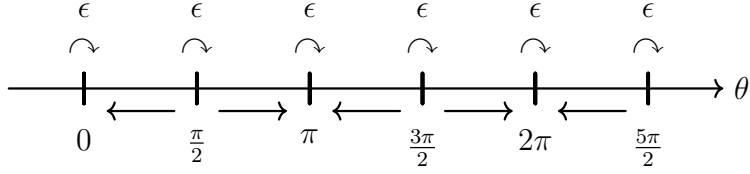

For the convenience of the reader, let us briefly recall from~\cite{DS} the intuitive description of the Pr\"ufer phase dynamics for $\epsilon\geq 0$. According to~\eqref{eq:factorization} and the group action property, it is useful to split the dynamics into two steps, first one induced by $D_{\kappa_{\sigma}}$ and the second by $R^\epsilon_{\sigma}$. Thus let us set for half-integers $n'=n-\frac{1}{2}$
\begin{equation}
\label{eq:Q-D-theta}
e_{\theta^\epsilon_n}
\;=\;
\frac{R^\epsilon_{n}\,
e_{\theta^\epsilon_{n'}}}{\|R^\epsilon_{n}\,
e_{\theta^\epsilon_{n'}}\|}
\,,
\qquad
e_{\theta^\epsilon_{n'}}
\;=\;
\frac{D_{n}\,
e_{\theta^\epsilon_{n-1}}}{\|D_{n}\,
e_{\theta^\epsilon_{n-1}}\|}
\,,
\end{equation}
where $D_n=D_{\kappa_n}$. The first step of the random dynamics induced by $D_{n}$ has fixed points at $\frac{\pi}{2}\,\mathbb{Z}$ and leaves each interval $(k\frac{\pi}{2},(k+1)\frac{\pi}{2})$ invariant. It will be explained and used below that in a logarithmic representation of the associated Dyson-Schmidt variables this dynamics becomes a random walk, with a supplementary drift in the unbalanced case. The second step induced by $R^\epsilon_{n}$ is a right shift (or clockwise rotation on the projected circle) by random angles of order $\epsilon$ because $a_n-|b_n|\geq C_1>0$ a.s. and
\begin{equation}
\label{eq:Q-S^1}
\theta_n^\epsilon
\;=\;
\theta_{n'}^\epsilon
\,+\,
\epsilon\big(a_n + b_n\cos(2\theta_{n'}^\epsilon)\big) 
\,+\, 
\mathcal{O}(\epsilon^2)
\,.
\end{equation}
Hence the combined dynamics passes through the fixed points $\frac{\pi}{2}\,\mathbb{Z}$ only in the increasing direction. This is illustrated in Figure~\ref{fig:theta-unbalanced}. Note that, in particular, the random dynamics passes in an alternating manner through a fixed point from $\pi \mathbb{Z}$ and one from $\pi(\frac{1}{2}+\mathbb{Z})$. Furthermore, one readily deduces a crucial order preserving property of the random dynamics, namely if one considers two further sequences $\widehat{\theta}^{\epsilon}_{n}$ and $\widetilde{\theta}^{\epsilon}_{n}$ constructed as in~\eqref{eq:Q-D-theta} with the same realization $\omega$, then 
\begin{equation}
\label{stat:order}
\widehat{\theta}^{\epsilon}_{0}
\;<\;
{\theta}^{\epsilon}_{0}
\;<\;
\widetilde{\theta}^{\epsilon}_{0}
\quad
\Longrightarrow
\quad
\widehat{\theta}^{\epsilon}_{n}
\;<\;
{\theta}^{\epsilon}_{n}
\;<\;
\widetilde{\theta}^{\epsilon}_{n}
\,,
\end{equation}
for all $n\in\frac{1}{2}\,\mathbb{Z}$. Based on this, it will be shown how to bound the dynamics above and below by two constructed processes. Then the rotation number in~\eqref{eq:IDOS-Birkhoff} can, via the elementary renewal theorem, be estimated by the inverse of their expected passage times through the intervals $(k\frac{\pi}{2},(k+1)\frac{\pi}{2})$.

%%%%%%%%%%%%%%%%%%%%%%%%%%%%%%%%%%
\section{Dyson-Schmidt variables and renewal processes}

The Dyson-Schmidt variable $x^\epsilon_n\in\mathbb{R}$ for $n\in\frac{1}{2}\mathbb{Z}$ associated to the Pr\"ufer phases is defined by
$$
x^\epsilon_n 
\;:=\; 
-\,
\cot(\theta^\epsilon_n)\,.
$$
This establishes an orientation preserving bijection of every interval $[k\pi,(k+1)\pi)$ with $k\in\mathbb{Z}$ to $\overline{\mathbb{R}}=\mathbb{R}\cup\{\infty\}$, with the central point $(k+\frac{1}{2})\pi$ being mapped to $0$. Then~\eqref{eq:Q-D-theta} becomes for $n\in\mathbb{Z}$ and $n'=n-\frac{1}{2}$
\begin{equation}
\label{eq:Q-D-x}
x^\epsilon_n
\;:=\;
-(R^\epsilon_{n}\cdot (-x^\epsilon_{n'}))
\:=\;
Q^\epsilon_{n}\cdot x^\epsilon_{n'}
\,,
\qquad
x^\epsilon_{n'}
\;:=\;
D_{n}\cdot x^\epsilon_{n-1}
\,,
\end{equation}
where the dot $\cdot$ denotes the standard M\"obius action and 
$$
Q^\epsilon_{n}
\;:=\;
\begin{pmatrix}
1 & 0 \\ 0 & -1
\end{pmatrix}
R^\epsilon_{n}
\begin{pmatrix}
1 & 0 \\ 0 & -1
\end{pmatrix}
\,,
$$
namely $Q^\epsilon_{n}$ is obtained from $R^\epsilon_{n}$ by flipping the signs on the off-diagonals. Due to the explicit forms of $Q^\epsilon_{n}$ and $D_{n}$, the action here becomes
\begin{equation}
\label{eq:Q-D-explicit}
x^\epsilon_n
\;=\;
\frac{(1+\epsilon^2\alpha^{\epsilon}_n)x^\epsilon_{n'}+(a_n{-b_n}-\epsilon\beta^{\epsilon}_n)\epsilon}{1+\epsilon^2\delta^{\epsilon}_n-(a_n{+b_n}+\epsilon\gamma^{\epsilon}_n)\epsilon x^\epsilon_{n'}}
\,,
\qquad
x^\epsilon_{n'}
\;=\;
\kappa_n^2 \, x^\epsilon_{n-1}
\,.
\end{equation}
Note that the interval $[k\pi,(k+1)\pi)$ of Pr\"ufer variables contains two fixed points $k\pi$ and $(k+\frac{1}{2})\pi$ of the dynamics generated by $D_\kappa$, so that one copy $\overline{\mathbb{R}}$ of the Dyson-Schmidt variable also contains two such fixed points $0$ and $\infty$. For the moment, only the half-axis $[0,\infty) \subset \overline{\mathbb{R}}$ will be considered. Below, it will be justified that this is indeed sufficient to show the results in Theorem~\ref{theo:SSH}. On this interval, it will be useful to take the logarithm. For this purpose, let us now state the main technical assumptions throughout the remainder of the paper.

\vspace{.2cm}

\noindent {\bf Hypothesis:} {\it The family $(\log(\kappa_n))_{n \geq 0}$ of random variables  is supposed to be independent and identically distributed with a non-trivial distribution in the sense that $\mathbb{P}\big(\{\log(\kappa_0) > 0\}\big)>0$.
This distribution is also assumed to have compact support, that is,
$$
C_0 
\;:=\;
\esssup \,|\log(\kappa_0)| \,\in\, (0,+\infty)\,,
$$
where the essential supremum is taken over (the suppressed index) $\sigma\in\Sigma$ w.r.t. the given distribution thereon. Furthermore the following constants are supposed to be positive and finite:}
$$
C_1
\;:=\;
\essinf\left(a_{\sigma}-|b_{\sigma}|\right)
\,,
\quad
C_2
\;:=\;
\esssup\big(a_{\sigma}+|b_{\sigma}|\big)
\,,
\quad
C_3
\;:=\;
\sup\limits_{|\epsilon|\leq 1}\esssup\|A_{\sigma}^{\epsilon}\|
\,.
$$
This implies, in particular, that one has at least for $\epsilon=0$
\begin{equation}
\label{eq:log}
\log(x^0_{n+1})
\;=\;
\log(x^0_{n})\;+\;2\,\log(\kappa_n)
\;=\;
\log(x^0_{n})\;+\;2\,C_0\,\chi_n
\,,
\end{equation}
where $\chi_n:=\frac{1}{C_0}\log(\kappa_n)$ is a random variable satisfying $-1 \leq \chi_n \leq 1$ almost surely. In the unbalanced case, $\mathbb{E}(\chi_n) <0$ while in the balanced $\mathbb{E}(\chi_n) = 0$. In both cases, $\log(x^0_{n})$ is a random walk on $\mathbb{R}$. It will be shown in the next two sections that in this logarithmic representation the random walk roughly has to go from $\log(\epsilon)$ to $-\log(\epsilon)$. The central limit theorem indicates that it takes of order of $\log(\epsilon)^2$ time steps to cross this distance of order $|\log(\epsilon)|$ in the balanced case, and turns out that in the unbalanced case an order of $\epsilon^{-\nu}$ time steps are needed. This provides an intuitive understanding for the behavior in Theorem~\ref{theo:SSH}. Controlling the $\epsilon$-dependent perturbations is quite delicate and the main technical endeavor of this work.

\vspace{.2cm}

The outcome will be bounds for the r.h.s. of~\eqref{eq:IDOS-Birkhoff}. In order to control the rotation number, it is useful to look at the order statistics of the following set of random variables
\begin{equation}
\label{stat:passages}
\big\lbrace N \in \mathbb{N} \,:\, x^\epsilon_{N-1} < 0 \leq x^\epsilon_{N} \mbox{ or }   x^\epsilon_{N} < 0 \leq x^\epsilon_{N-1} \big\rbrace\,,
\end{equation}
which will be denoted by the random increasing times $N^\epsilon_{(1)}<N^\epsilon_{(2)}<N^\epsilon_{(3)}<...$. These are the random passage times over the two points $0$ and $\infty$ (which are fixed points of the action induced by $D^0_{n}$, so without $\epsilon$-perturbation). Recall that the two conditions in~\eqref{stat:passages} are realized in an alternating manner. For sake of concreteness, let us fix the initial condition $x_0^\epsilon\in (-\infty,0)\cup\{\infty\}$ such that for all $k$ one has $x^\epsilon_{N^\epsilon_{(2k-1)}} < 0 \leq x^\epsilon_{N^\epsilon_{(2k-1)}-1}$ and $x^\epsilon_{N^\epsilon_{(2k)}-1} < 0 \leq x^\epsilon_{N^\epsilon_{(2k)}}$.  The (random) differences $N^\epsilon_{(k+1)}-N^\epsilon_{(k)}$ are the durations of the passages of $x$ through the intervals $[0,+\infty)$ and $\overline{\mathbb{R}} \backslash [0,+\infty)$. Clearly, these quantities depend on the precise value of the initial condition $x_0^\epsilon$ and therefore they are not identically distributed (nor independent). To circumvent this difficulty, two families of random dynamical processes $\widehat{x}^\epsilon_k=(\widehat{x}^\epsilon_{k,n})_{n\geq 0}$ and $\widetilde{x}^\epsilon_k=(\widetilde{x}^\epsilon_{k,n})_{n\geq 0}$ on $[0,\infty)$ will be constructed for all $k \in \mathbb{N}$ in the two following sections, providing lower and upper bounds on the original process respectively. It will be imposed that $(\widehat{x}^\epsilon_{2k-1})_{k\in\mathbb{N}}$, $(\widehat{x}^\epsilon_{2k})_{k\in\mathbb{N}}$, $(\widetilde{x}^\epsilon_{2k-1})_{k\in\mathbb{N}}$ and $(\widetilde{x}^\epsilon_{2k})_{k\in\mathbb{N}}$ are all families of nonnegative i.i.d. random variables. Note that these processes will not exactly correspond to the notations $\widehat{\theta}^\epsilon_n$ and $\widetilde{\theta}^\epsilon_n$ in~\eqref{stat:order}. They will obey almost surely  that for all $k \in \mathbb{N}$ and then all $n \in \lbrace 0,1,\dots,N^\epsilon_{(k+1)} - N^\epsilon_{(k)}-1\rbrace$
\begin{equation}
\label{stat:slower-faster}
\widehat{x}_{k,n}^\epsilon\,\leq\, x^\epsilon_{N^\epsilon_{(k)} + n}\, \leq\, \widetilde{x}_{k,n}^\epsilon
\qquad\mbox{or}\qquad
\widehat{x}_{k,n}^\epsilon\,\leq\, -\big(x^\epsilon_{N^\epsilon_{(k)} + n}\big)^{-1}\, \leq\, \widetilde{x}_{k,n}^\epsilon
\end{equation}
and furthermore, for the next step
\begin{equation}
\label{stat:slower-faster-bis}
\widetilde{x}^\epsilon_{k,N^\epsilon_{(k+1)} - N^\epsilon_{(k)}}
\;=\;
\infty\,.
\end{equation}
Note that the left condition in~\eqref{stat:slower-faster} always applies during passages through $(0,\infty)$ and the right condition for passages through $(-\infty,0)$. Moreover, the constructed comparison processes are only constrained  on the first $N^\epsilon_{(k+1)} - N^\epsilon_{(k)}$ times. Associated to these two families of processes, there are now two families of random passage times
\begin{equation}
\label{eq:T}
\widehat{T}^\epsilon_k
\;:=\;
\inf\lbrace n \in \mathbb{N}_0 \,:\, \widehat{x}^\epsilon_{k,n} = \infty\rbrace\,,
\qquad 
\widetilde{T}^\epsilon_k
\;:=\;
\inf\lbrace n \in \mathbb{N}_0 \,:\, \widetilde{x}^\epsilon_{k,n} = \infty\rbrace\,.
\end{equation}
Then~\eqref{stat:slower-faster} and~\eqref{stat:slower-faster-bis} imply that a.s. $\widetilde{T}^\epsilon_k \leq N^\epsilon_{(k+1)} - N^\epsilon_{(k)} \leq \widehat{T}^\epsilon_k$. Furthermore, by construction the families $(\widehat{T}^\epsilon_{2k-1})_{k\in\mathbb{N}}$, $(\widehat{T}^\epsilon_{2k})_{k\in\mathbb{N}}$, $(\widetilde{T}^\epsilon_{2k-1})_{k\in\mathbb{N}}$ and $(\widetilde{T}^\epsilon_{2k})_{k\in\mathbb{N}}$ are i.i.d. random variables. As then $(\widehat{T}^\epsilon_{2k-1}+\widehat{T}^\epsilon_{2k})_{k\in\mathbb{N}}$ and $(\widetilde{T}^\epsilon_{2k-1}+\widetilde{T}^\epsilon_{2k})_{k\in\mathbb{N}}$ are interarrival times~\cite{GS}, both families specify a renewal process, given by
$$
\widehat{P}^\epsilon_N 
\,:=\,
\max\Big\lbrace K \in \mathbb{N} \,:\sum_{k=1}^K (\widehat{T}^\epsilon_{2k-1}+\widehat{T}^\epsilon_{2k}) \leq N\Big\rbrace\,,
\;\;
\widetilde{P}^\epsilon_N 
\,:=\, 
\max\Big\lbrace K \in \mathbb{N} \,: \sum_{k=1}^K (\widetilde{T}^\epsilon_{2k-1}+\widetilde{T}^\epsilon_{2k}) \leq N\Big\rbrace\,.
$$
These random variables can be interpreted as the number of times the slower or faster process has passed through $\overline{\mathbb{R}}$ up to the time $N$. Each such passage corresponds to a passage of the Pr\"ufer variables through $[k\pi,(k+1)\pi)$. Thus it follows that
$$
\widehat{P}^\epsilon_N - 1
\;\leq\; 
\frac{\theta^\epsilon_N}{\pi} 
\;\leq\; 
\widetilde{P}^\epsilon_N + 1
$$
a.s. for $N \in \mathbb{N}_0$ and $\theta^\epsilon_0 \in \left[-\frac{\pi}{2},\frac{\pi}{2}\right)$. Finally, the elementary renewal theorem~\cite{GS} yields
\begin{equation}
\label{ineq:slower-faster}
\frac{1}{\mathbb{E}(\widehat{T}_1^\epsilon)+\mathbb{E}(\widehat{T}_2^\epsilon)} 
\;=\; 
\lim_{N \to \infty} \frac{\widehat{P}^\epsilon_N}{N} 
\;\leq\; 
\lim_{N \to \infty} \frac{1}{N}\frac{\mathbb{E}(\theta^\epsilon_N)}{\pi} 
\;\leq\; 
\lim_{N \to \infty} \frac{\widetilde{P}^\epsilon_N}{N} 
\;=\; 
\frac{1}{\mathbb{E}(\widetilde{T}_1^\epsilon)+\mathbb{E}(\widetilde{T}_2^\epsilon)}\,.
\end{equation}
These bounds hold for both the unbalanced and the balanced case.

\vspace{.2cm}

Let us now first address the unbalanced case. The opposite directions of the drifts in Figure~\ref{fig:theta-unbalanced} clearly show that $\mathbb{E}(\widehat{T}_1^\epsilon) \geq \mathbb{E}(\widehat{T}_2^\epsilon)$. Thus the inverse of $2\,\mathbb{E}(\widehat{T}_1^\epsilon)$ provides a lower bound on the l.h.s. of \eqref{ineq:slower-faster}, while the r.h.s. can simply be estimated by the inverse of $\mathbb{E}(\widetilde{T}_1^\epsilon)$. These rough estimates are actually superfluous, since the contributions of $\mathbb{E}(\widehat{T}_1^\epsilon)$ and $\mathbb{E}(\widetilde{T}_1^\epsilon)$ turn out to dominate those of $\mathbb{E}(\widehat{T}_2^\epsilon)$ and $\mathbb{E}(\widetilde{T}_2^\epsilon)$ for $\epsilon$ going to $0$. For this reason, passages through $(-\infty,0)$ need not to be taken into account, so restricting the following to the case $k=1$  is sufficient (this corresponds to the first passage with positive $x$ variables). The proof of the following result will be provided in the next two sections. Combining it with~\eqref{ineq:slower-faster} directly implies the claim \eqref{eq:pseudo} in Theorem~\ref{theo:SSH}. 

%%%%%%%%%%%%%%%%%%%%%%%%%%%%%%%%%%%%%%%%%%%%%%%%
\begin{proposition}
\label{prop:T}
Given a family of random matrices of the form~\eqref{eq:factorization}, suppose that the above Hypothesis holds and $\mathbb{E}(\log(\kappa_0)) < 0$. Then there exists a unique positive solution $\nu>0$ of $\mathbb{E}(\kappa_0^{\nu}) = 1$. Moreover, there exist constants $C_-, C_+ \in (0,\infty)$ such that for all $\widetilde{\nu} < \nu$ there exists some $\epsilon_0$ with then for all $\epsilon \in (0,\epsilon_0)$
$$
\frac{1}{\mathbb{E}(\widehat{T}_1^\epsilon)} 
\;\geq\;
C_-\epsilon^{\nu}\,\big(1+\mathcal{O}(\epsilon^{\nu}|\log(\epsilon)|)\big)
\,,
\qquad
\frac{1}{\mathbb{E}(\widetilde{T}_1^\epsilon)} 
\;\leq\;
C_+\epsilon^{\widetilde{\nu}}\,\big(1+\mathcal{O}(\epsilon^{\widetilde{\nu}}|\log(\epsilon)|)\big)
\,.
$$
\end{proposition}
%%%%%%%%%%%%%%%%%%%%%%%%%%%%%%%%%%%%%%%%%%%%%%%%

In the balanced case, $\mathbb{E}(\widehat{T}_k^\epsilon)=\mathbb{E}(\widetilde{T}_k^\epsilon)$ to lowest order. This value is independent of $k$ and can be computed, as the next result shows. Together  with \eqref{ineq:slower-faster} this shows
\eqref{ineq:log-div} in Theorem~\ref{theo:SSH}.

%%%%%%%%%%%%%%%%%%%%%%%%%%%%%%%%%%%%%%%%%%%%%%%%
\begin{proposition}
\label{prop:T-balanced}
Given a family of random matrices of the form~\eqref{eq:factorization}, suppose that the above Hypothesis holds and $\mathbb{E}(\log(\kappa_0)) = 0$. Then for all $k$ it holds that
$$
\frac{1}{\mathbb{E}(\widehat{T}_k^\epsilon)} 
\;=\;
\frac{\mathbb{E}(\log(\kappa_0)^2)}{\log(\epsilon)^2}\,\big(1+\mathcal{O}(|\log(\epsilon)|^{-1})\big) 
\,,
\qquad
\frac{1}{\mathbb{E}(\widetilde{T}_k^\epsilon)} 
\;=\;
\frac{\mathbb{E}(\log(\kappa_0)^2)}{\log(\epsilon)^2}\,\big(1+\mathcal{O}(|\log(\epsilon)|^{-1})\big) 
\,.
$$
\end{proposition}
%%%%%%%%%%%%%%%%%%%%%%%%%%%%%%%%%%%%%%%%%%%%%%%%

%%%%%%%%%%%%%%%%%%%%%%%%%%%%%%%%%%%%%%%%%%%
\section{Lower bound on the rotation number}
\label{sec:lower}

The first task of this section is to construct the slower comparison process satisfying the first inequality of~\eqref{stat:slower-faster} for $k=1$. To improve readability from this point on, we will suppress the indices $\epsilon$, $\sigma$, $n$, etc., as long as no confusion can arise. Let us start by providing some basic properties of the dynamics, such as the following observation.

%%%%%%%%%%%%%%%%%%%%%%%%%%%%%%%%%%%%%%%%%%%%%%%%
\begin{lemma}
\label{lemma:Q-lower}
For each realization and $\epsilon$ small enough, $x \in [0,\infty)$ and $Q \cdot x \geq 0$ imply $Q \cdot x \geq x$.
\end{lemma}
%%%%%%%%%%%%%%%%%%%%%%%%%%%%%%%%%%%%%%%%%%%%%%%%

\noindent {\bf Proof.} The main Hypothesis implies for $x \in [0,1]$ the estimates
$$
Q \cdot x
\;=\; 
\tfrac{(1+\epsilon^2\alpha)x + (a-b-\epsilon\beta)\epsilon}{1+\epsilon^2\delta - (a+b+\epsilon\gamma)\epsilon x}
\,\geq\,
x\,\tfrac{1+\epsilon^2\alpha + (a-b-\epsilon\beta)\frac{\epsilon}{x}}{1+\epsilon^2\delta} 
\,\geq \,
x\,\tfrac{1+C_1\,\epsilon-2\,C_3\,\epsilon^2}{1+C_3\,\epsilon^2}\,, 
$$
while for $x \in (1,\infty)$
$$
Q \cdot x
\,\geq\,
x\,\tfrac{1+\epsilon^2\alpha}{1+\epsilon^2\delta - (a+b+\epsilon\gamma)\epsilon x}
\,\geq\,
x\,\tfrac{1+\epsilon^2\alpha}{1+\epsilon^2\delta - (a+b+\epsilon\gamma)\epsilon }
\,\geq\, 
x\,\tfrac{1-C_3\,\epsilon^2}{1-C_1\,\epsilon+2\,C_3\,\epsilon^2}
\,.
$$
In both cases the statement directly follows. (Note that the lemma can also be deduced from~\eqref{eq:Q-S^1}.)
\hfill $\square$

\vspace{.2cm}

The following lemma states more properties of the given action, and relies on the quantities
$$
\widehat{x}_-
\;:=\;
\tfrac{C_1\,\epsilon}{2}
\,,
\qquad
\widehat{x}_c
\;:=\;
\tfrac{C_1\epsilon}{2}(e^{-2C_0}+1)
\,,
\qquad
\widehat{x}_+
\;:=\;
\tfrac{2\,e^{2C_0}}{C_1\epsilon}
\,.
$$
These points and the lemma itself are graphically illustrated in the left part of Figure~\ref{fig:slower}.

\begin{figure}
	\begin{center}
		\begin{tikzpicture}[line join = round, line cap = round]
		\coordinate (a) at (0.0,0.3);
		\coordinate (b) at (0.0,-0.2);
		\coordinate (bb) at (0.0,-0.8);
		\coordinate (r) at (0.2,0.0);
		\coordinate (l0) at (-6.5,0);
		\tick{(l0)};
		\coordinate (l0a) at ($(l0) + (a)$);
		\coordinate[label=below:{$0$}] (l0b) at ($(l0) + (b)$);
		\coordinate (l1) at (-5.5,0);
		\tick{(l1)};
		\coordinate (l1a) at ($(l1) + (a)$);
		\coordinate (l1ar) at ($(l1) + (a) + (r)$);
		\draw[->] (l0a) to[out=90, in=105, looseness=1.5] node[midway,above,inner sep=4pt] {\eqref{stat:slower-start}} (l1ar);
		\coordinate[label=below:{$\widehat{x}_-$}] (l1b) at ($(l1) + (b)$);
		\coordinate (l2) at (-4.5,0);
		\tick{(l2)};
		\coordinate (l2ar) at ($(l2) + (a) + (r)$);
		\draw[->] (l1a) to[out=90, in=105, looseness=1.5] node[midway,above,inner sep=4pt] {\eqref{stat:slower-between}} (l2ar);
		\coordinate[label=below:{$\widehat{x}_c$}] (l2b) at ($(l2) + (b)$);
		\coordinate (l3) at (-4,0);
		\coordinate (l4) at (-3,0);
		\coordinate (l5) at (-2.5,0);
		\tick{(l5)};
		\coordinate (l5a) at ($(l5) + (a)$);
		\coordinate[label=below:{$\widehat{x}_+$}] (l5b) at ($(l5) + (b)$);
		\coordinate (lr) at (-1.5,0);
		\coordinate[label=left:{$\widehat{x}$}] (lrr) at (-1,0);
		\coordinate (lrra) at ($(lrr) + (a)$);
		\draw[->] (l5a) to[out=90, in=120, looseness=1.5] node[midway,above,inner sep=4pt] {\eqref{stat:slower-end}} (lrra);
		\draw [-,color=black,line width=0.3mm] (l0)--(l3);
		\draw [-,color=black,dotted,line width=0.3mm] (l3)--(l4);
		\draw [->,color=black,line width=0.3mm] (l4) -- (lr);
		\coordinate (ml) at (-0.5,-0.5);
		\coordinate (mr) at (0.5,-0.5);
		\draw[->] (ml) to[bend right] node[midway,above,inner sep=4pt] {$\widehat{f}$} (mr);
		\coordinate (rl) at (1.5,0);
		\coordinate (r1) at (2,0);
		\tick{(r1)};
		\coordinate[label=below:{$\widehat{y}_-$}] (r1b) at ($(r1) + (b)$);
		\coordinate (r2) at (3,0);
		\tick{(r2)};
		\coordinate[label=below:{$0$}] (r2b) at ($(r2) + (b)$);
		\coordinate (r3) at (3.5,0);
		\coordinate (r4) at (4.5,0);
		\coordinate (r5) at (5,0);
		\tick{(r5)};
		\coordinate[label=below:{$\widehat{y}_+$}] (r5b) at ($(r5) + (b)$);
		\coordinate[label=right:{$\widehat{y}$}] (rr) at (5.5,0);
		\draw [-,color=black,line width=0.3mm] (rl)--(r3);
		\draw [-,color=black,dotted,line width=0.3mm] (r3)--(r4);
		\draw [->,color=black,line width=0.3mm] (r4) -- (rr);
		\end{tikzpicture}
		\caption{\it The arrows on the left part illustrate properties of the original Dyson-Schmidt dynamics on $(0,\infty)$ as stated in {\rm Lemma~\ref{lemma:slower}}. The right part illustrates the notations after the logarithmic transformation $\widehat{f}$ to $\mathbb{R}$.}
		\label{fig:slower}
	\end{center}
\end{figure}
	
%%%%%%%%%%%%%%%%%%%%%%%%%%%%%%%%%%%%%%%
\begin{lemma}
\label{lemma:slower}
For each realization, one has
\begin{alignat}{3}
& x \,\in\, [0,\infty)  &\qquad \Longrightarrow \qquad & Q \cdot (D \cdot x) \,\notin\, [0,\widehat{x}_-)\,,
\label{stat:slower-start}\\
& x \,\in\, [\widehat{x}_-,\infty)  &\qquad \Longrightarrow \qquad  &  Q \cdot (D \cdot x) \,\notin\, [0,\widehat{x}_c)\,,
\label{stat:slower-between}\\
& Q \cdot (D \cdot x) \,\in\, [0,\infty)  &\qquad \Longrightarrow \qquad  & x \,\notin\, [\widehat{x}_+,\infty)\,.
\label{stat:slower-end}
\end{alignat}
\end{lemma}
%%%%%%%%%%%%%%%%%%%%%%%%%%%%%%%%%%%%%%%

\noindent {\bf Proof.} For~\eqref{stat:slower-start}, first note that $x \in [0,\infty)$ implies $D \cdot x \in [0,\infty)$. By combining~\eqref{eq:Q-S^1} and the order-preserving property~\eqref{stat:order}, then nonnegative $Q \cdot (D \cdot x)$ obey, due to~\eqref{eq:Q-D-explicit} and the Hypothesis,
$$
Q \cdot (D \cdot x) 
\;\geq\;
Q \cdot 0 
\;=\; 
\tfrac{(a-b-\epsilon\beta)\epsilon}{1+\epsilon^2\delta} 
\;\geq\; 
\tfrac{(C_1-C_3\epsilon)\epsilon}{1+C_3\epsilon^2} 
\;\geq\; 
\tfrac{C_1\,\epsilon}{2}
\;=\;
\widehat{x}_-
\,.
$$
For the proof of~\eqref{stat:slower-between} let us use that, if $x \in [\widehat{x}_-,\infty)$, then clearly $D \cdot x \geq e^{-2C_0}\widehat{x}_-$. Similarly to the proof of~\eqref{stat:slower-start}, nonnegative $Q \cdot (D \cdot x)$ then obey
$$
Q \cdot (D \cdot x) 
\;\geq\; 
Q \cdot \tfrac{e^{-2C_0}C_1\epsilon}{2} 
\;\geq\; 
\tfrac{(1-C_3\epsilon^2)\frac{e^{-2C_0}C_1\epsilon}{2} + (C_1-C_3\epsilon)\epsilon}{1+C_3\epsilon^2 - (C_1-C_3\epsilon)\frac{e^{-2C_0}C_1\epsilon^2}{2}} 
\;\geq\; 
(e^{-2C_0} + 1)\tfrac{C_1\epsilon}{2}
\;=\;\widehat{x}_c
\,.
$$
Finally let us verify~\eqref{stat:slower-end} by contraposition. If $x \in [\widehat{x}_+,\infty)$, then clearly $D \cdot x \geq e^{-2C_0}\widehat{x}_+=\tfrac{2}{C_1\epsilon}$. Then the order-preserving property~\eqref{stat:order} implies that $Q \cdot (D \cdot x) \notin [0,\infty)$, since as in the proof of~\eqref{stat:slower-start} it holds that 
$$
0 
\;>\; 
Q \cdot \infty 
\;\geq\;
Q \cdot (D \cdot x) 
\;\geq\; 
Q \cdot \tfrac{2}{C_1\epsilon} 
\;\geq\; 
\tfrac{(1-C_3\epsilon^2)\frac{2}{C_1\epsilon} + (C_1-C_3\epsilon)\epsilon}{1+C_3\epsilon^2 - (C_1-C_3\epsilon)\frac{2}{C_1}}
\,,
$$ 
which is also negative.
\hfill $\square$

\vspace{.2cm}

Now a new process $\widehat{x}=(\widehat{x}_n)_{n\geq 0}$ is constructed by setting $\widehat{x}_0 = 0$, $\widehat{x}_1=\widehat{x}_-$ and for $n\geq 1$
$$
\widehat{x}_{n+1} 
\;=\; 
\begin{cases}
\widehat{x}_c \,,&\text{ if } \widehat{x}_n \leq \widehat{x}_-\,,\\
D_n \cdot \widehat{x}_n \;,&\text{ if } \widehat{x}_n \in (\widehat{x}_-,\widehat{x}_+)\,,
\\
\infty\,, &\text{ else, so if } \widehat{x}_n \geq  \widehat{x}_+\,.
\end{cases}
$$
Comparing with~\eqref{eq:Q-D-x}, the main case $\widehat{x}_{n+1}=D_n \cdot \widehat{x}_n$ of this process merely omits the action of $Q_n$ for $n\geq 2$. Let us now argue why this process satisfies the first inequality in~\eqref{stat:slower-faster}. Indeed, omitting the action of $Q_n$ slows the process down because of the order-preserving property~\eqref{stat:order} and Lemma~\ref{lemma:Q-lower}. Carefully analyzing the first case in the definition of $\widehat{x}_{n+1}$ in combination with~\eqref{stat:slower-start} and~\eqref{stat:slower-between} shows that a.s. $\widehat{x}_n \leq x_{N_{(1)} + n}$ for all $n \in \lbrace 0,1,\dots,N_{(2)} - 1 - N_{(1)}\rbrace$, that is, as long as $x_{N_{(1)} + n} \in [0,\infty)$. Moreover, by~\eqref{stat:slower-end} it is impossible that $\widehat{x}_n = \infty$ for $n \in \lbrace 3,4,\dots,N_{(2)} - 1 - N_{(1)}\rbrace$, as this would imply $x_{N_{(1)} + n-1} < \widehat{x}_+ \leq \widehat{x}_{n-1}$. Conversely, as a.s. $\widehat{x}_{\widehat{T}_1} = \infty$, then indeed $N_{(2)} - N_{(1)} \leq \widehat{T}_1$. 

\vspace{.2cm}

Now let us come to the second task, namely analyzing the $\epsilon$-dependence of $\big(\mathbb{E}(\widehat{T}_1)\big)^{-1}$ and thereby proving the first statement of Proposition~\ref{prop:T}. Similar as in~\eqref{eq:log}, it will be advantageous to pass to a shifted logarithm of the Dyson-Schmidt variables, via the map $\widehat{f}:(0,\infty)\to \mathbb{R}$ given by 
$$
\widehat{f}(x)
\;:=\;
\frac{1}{2C_0}\,\log\Big(\frac{x}{\widehat{x}_c}\Big)
\,.
$$
By construction, $\widehat{f}(\widehat{x}_c) = 0$. Furthermore, for $n$ such that $\widehat{x}_{n+2}<\infty$, let us introduce 
$$
\widehat{y}_n
\;:=\;
\widehat{f}(\widehat{x}_{n+2})
\,,
\qquad
\widehat{y}_-\;:=\;\widehat{f}(\widehat{x}_-)
\,,
\qquad
\widehat{y}_+\;:=\;\widehat{f}(\widehat{x}_+)
\,,
$$
and the stopping time
$$
\widehat{T}_{-,+}
\;:=\;
\inf\big\{ n \in \mathbb{N} \,:\, 
\widehat{y}_n\notin (\widehat{y}_-,\widehat{y}_+) 
\big\}\,.
$$
Again these quantities are illustrated in Figure~\ref{fig:slower}. As long as $n \leq \widehat{T}_{-,+}$, it holds that
\begin{equation}
\label{eq:y-slower}
\widehat{y}_n
\;=\; 
\tfrac{1}{2C_0}\,\log\Big(\frac{\widehat{x}_{n+2}}{\widehat{x}_c}\Big)
\;=\; 
\tfrac{1}{2C_0}\log\Big(\frac{D^n \cdot \widehat{x}_2}{\widehat{x}_c}\Big)
\;=\; 
\tfrac{1}{2C_0}\log\Big(\prod_{j=N_{(1)}+1}^{N_{(1)}+n} \kappa_j^2\Big) 
\;=\; 
\sum_{j=N_{(1)}+1}^{N_{(1)}+n} \chi_j\,,
\end{equation}
namely $ \widehat{y}_n $ is a random walk with a drift in the negative direction starting at $\widehat{y}_0=0$. The following two lemmata recollect properties about these newly introduced quantities.

%%%%%%%%%%%%%%%%%%%%%%%%%%%%%%%%%%%%%%%
\begin{lemma}
\label{lemma:slower-y-lims}
$\widehat{y}_- \in (-\infty,0)$ is independent of $\epsilon$ and $\lim_{\epsilon \to 0} \frac{\widehat{y}_+}{-\log(\epsilon)} = \frac{1}{C_0}$.
\end{lemma}
%%%%%%%%%%%%%%%%%%%%%%%%%%%%%%%%%%%%%%%

\noindent\textbf{Proof.} The explicit expressions
$$
\widehat{y}_- \,=\, -\tfrac{1}{2\,C_0}\,\log(1+e^{-2\,C_0})\;,
\qquad
\widehat{y}_+ \,=\, \tfrac{1}{2\,C_0}\big(2\log\big(\tfrac{2}{C_1\,\epsilon}\big) -\log(1+e^{-2\,C_0})\big) +1
\,,
$$
immediately imply the claims.
\hfill $\square$

%%%%%%%%%%%%%%%%%%%%%%%%%%%%%%%%%%%%%%%
\begin{lemma}
\label{lemma:E-T-slower}
$\mathbb{E}(\widehat{T}_{-,+} ) < +\infty$.
\end{lemma}
%%%%%%%%%%%%%%%%%%%%%%%%%%%%%%%%%%%%%%%

\noindent\textbf{Proof.}
Since the cumulative distribution function of $\chi$ is right-continuous and $\mathbb{P}(\{\chi > 0\}) > 0$, there exists some $\ell \in (0,1]$ such that $ \widehat{p}:=\mathbb{P}(\{\chi \geq \ell\})$ satisfies  $\widehat{p} > 0$. Denoting $\widehat{E} := \lceil\frac{\widehat{y}_+-\widehat{y}_-}{\ell}\rceil$ and introducing the random variable 
$$
\widehat{N}
\;:=\;
\min\lbrace n \in \mathbb{N} \,:\, \chi_{(n-1)\widehat{E} + 1} \geq \ell\;,\; \chi_{(n-1)\widehat{E} + 2} \geq \ell\;, \dots,\; \chi_{n\widehat{E}} \geq \ell \rbrace
\,,
$$
the latter then is geometrically distributed with success probability $\widehat{p}^{\widehat{E}}$. In particular, one has $\mathbb{E}(\widehat{N}) < \infty$. Moreover, $\widehat{T}_{-,+} < \widehat{E}\,\widehat{N}$ a.s. by construction, so $\mathbb{E}(\widehat{T}_{-,+}) < \widehat{E}\,\mathbb{E}(\widehat{N}) < \infty$.
\hfill $\square$

\vspace{.2cm}

In order to connect the two stopping times $\widehat{T}_{-,+}$ and $\widehat{T}_1$, one further random variable will be introduced. Suppose that $\widehat{T}_{-,+} = m$ for some $m \in \mathbb{N}$, and $\widehat{y}_{\widehat{T}_{-,+}}\leq\widehat{y}_-$, then let us introduce the at $m+1$ reinitialized stopping time as in~\eqref{eq:T} by
$$
\widehat{T}^{(m)}_1 
\;:=\; 
\inf\big\{ n \in \mathbb{N} \,:\, \widehat{x}_{m+1+n} = \infty \big\}
\,.
$$
It then clearly follows that $\widehat{T}_1 = m + 1 + \widehat{T}^{(m)}_1$, provided that $\widehat{T}_{-,+} = m$ and $\widehat{y}_{m}\leq\widehat{y}_-$. Now the Markov property allows to compute the conditional expectations
$$
\mathbb{E}\big(\widehat{T}^{(m)}_1 \,\big|\, \widehat{y}_{m} \leq \widehat{y}_-\,,\;\; \widehat{T}_{-,+}=m\big) 
\;=\; 
\mathbb{E}(\widehat{T}_1)
\,,\qquad
\mathbb{E}\big(\widehat{T}_1 \,\big|\, \widehat{y}_{\widehat{T}_{-,+}} \geq \widehat{y}_+\big)
\;=\; 
\mathbb{E}\big(\widehat{T}_{-,+} + 3 \,\big|\, \widehat{y}_{\widehat{T}_{-,+}} \geq \widehat{y}_+\big)
\,,
$$
where the $3$ stems from an index shift by $2$ when the process is started and $1$ additional step at the end. As by construction $\mathbb{P}\big(\{\widehat{y}_{\widehat{T}_{-,+}}\in(\widehat{y}_-,\widehat{y}_+)\}\big)=0$ and as by Lemma~\ref{lemma:E-T-slower} one has $\widehat{T}_{-,+}<\infty$ a.s., it follows that
\begin{align*}
\begin{aligned}
	\mathbb{E}(\widehat{T}_1) 
	&\;=\; \mathbb{E}\big(\widehat{T}_1 \,\big|\, \widehat{y}_{\widehat{T}_{-,+}} \geq \widehat{y}_+\big)\,\mathbb{P}\big(\big\{\widehat{y}_{\widehat{T}_{-,+}} \geq \widehat{y}_+\big\}\big) 
	\\
	&\qquad + \sum_{m=0}^{\infty} \mathbb{E}\big(\widehat{T}_1 \,\big|\, \widehat{y}_m \leq \widehat{y}_-\,, \widehat{T}_{-,+}  = m\big)\,\mathbb{P}\big(\big\{\widehat{y}_m \leq \widehat{y}_-\,, \widehat{T}_{-,+}  = m\big\}\big)\\
	&\;=\; \mathbb{P}\big(\big\{\widehat{y}_{\widehat{T}_{-,+}} \geq \widehat{y}_+\big\}\big)\,\mathbb{E}\big(\widehat{T}_1 \,\big|\, \widehat{y}_{\widehat{T}_{-,+}} \geq \widehat{y}_+\big) 
	\\
	&\qquad + \sum_{m=0}^{\infty} \mathbb{P}\big(\big\{\widehat{y}_m \leq \widehat{y}_-\,, \widehat{T}_{-,+}  = m\big\}\big)\mathbb{E}\big(m + 1 + \widehat{T}^{(m)}_1 \,\big|\, \widehat{y}_{m} \leq \widehat{y}_-\,, \widehat{T}_{-,+}  = m\big)
	\\
	&\;= \;\mathbb{P}\big(\big\{\widehat{y}_{\widehat{T}_{-,+}} \geq \widehat{y}_+\big\}\big)\,\mathbb{E}\big(\widehat{T}_{-,+} + 3 \,\big|\, \widehat{y}_{\widehat{T}_{-,+}} \geq \widehat{y}_+\big)
	\\
	&\qquad + \sum_{m=0}^{\infty} \mathbb{P}\big(\big\{\widehat{y}_m \leq \widehat{y}_-\,, \widehat{T}_{-,+}  = m\big\}\big)\left(\mathbb{E}\big(\widehat{T}_{-,+} \,\big|\, \widehat{y}_{\widehat{T}_{-,+}} \leq \widehat{y}_-\,, \widehat{T}_{-,+}  = m\big) + 1 + \mathbb{E}(\widehat{T}_1)\right)
	\\
	&\;=\; \mathbb{E}(\widehat{T}_{-,+} )\, +\, 3\,\mathbb{P}\big(\big\{\widehat{y}_{\widehat{T}_{-,+}} \geq \widehat{y}_+\big\}\big)\, +\, \left(1-\mathbb{P}\big(\big\{\widehat{y}_{\widehat{T}_{-,+}} \geq \widehat{y}_+\big\}\big)\right)\left(1 + \mathbb{E}(\widehat{T}_1)\right)\,,
\end{aligned}
\end{align*}
which is equivalent to
\begin{equation}
\label{eq:slower}
\big(\mathbb{E}(\widehat{T}_1)\big)^{-1}
\;=\;
\left[\frac{\mathbb{E}(\widehat{T}_{-,+}) + 1}{\mathbb{P}\big(\big\{\widehat{y}_{\widehat{T}_{-,+}} \geq \widehat{y}_+\big\}\big)}\, +\, 2\right]^{-1}\,.
\end{equation}
It now remains to compute the probability and the expectation on the r.h.s. of~\eqref{eq:slower}. This will essentially follow from the optional stopping theorem. It is convenient to define the quantities
\begin{align*}
\widehat{y}'_-
&\;:=\; \mathbb{E}\big(\widehat{y}_{\widehat{T}_{-,+}} \,\big|\, \widehat{y}_{\widehat{T}_{-,+}} \leq \widehat{y}_-\big)\,,
&\qquad&
\widehat{y}''_-
\;:=\;
\tfrac{1}{C_0\nu}\log\Big(\mathbb{E}\big(e^{C_0\nu\widehat{y}_{\widehat{T}_{-,+}}} \,\big|\, \widehat{y}_{\widehat{T}_{-,+}} \leq \widehat{y}_-\big)\Big)\,,\\
\widehat{y}'_+
&\;:=\;
\mathbb{E}\big(\widehat{y}_{\widehat{T}_{-,+}} \,\big|\, \widehat{y}_{\widehat{T}_{-,+}} \geq \widehat{y}_+\big)\,,
&\qquad&
\widehat{y}''_+
\;:=\;
\tfrac{1}{C_0\nu}\log\Big(\mathbb{E}\big(e^{C_0\nu\widehat{y}_{\widehat{T}_{-,+}}} \,\big|\, \widehat{y}_{\widehat{T}_{-,+}} \geq \widehat{y}_+\big)\Big)\,,
\end{align*}
in which $\widehat{y}'_-, \widehat{y}''_- \in \left[\widehat{y}_--1,\widehat{y}_-\right]$ and $\widehat{y}'_+, \widehat{y}''_+ \in \left[\widehat{y}_+,\widehat{y}_++1\right]$. Now~\eqref{eq:y-slower} implies that $\widehat{y}_n - n\mathbb{E}(\chi)$ is a martingale. As $|\chi| \leq 1$ a.s., its increments are a.s. bounded, namely more precisely $|\widehat{y}_{n+1} - (n+1)\mathbb{E}(\chi) - \widehat{y}_n + n\mathbb{E}(\chi)| \leq 2$. Then with $\mathbb{E}(\widehat{T}_{-,+}) < \infty$ from Lemma~\ref{lemma:E-T-slower}, one can use the optional stopping theorem to find $0 = \mathbb{E}(\widehat{y}_0 - 0 \cdot \mathbb{E}(\chi)) = \mathbb{E}(\widehat{y}_{\widehat{T}_{-,+}} - \widehat{T}_{-,+} \cdot \mathbb{E}(\chi))$, or
\begin{equation}
\label{eq:slower-bis}
\mathbb{E}(\widehat{T}_{-,+}) 
\;=\; 
\frac{\mathbb{E}\big(\widehat{y}_{\widehat{T}_{-,+}}\big)}{\mathbb{E}(\chi)} 
\;=\; 
\frac{\widehat{y}'_-\big(1 - \mathbb{P}\big(\{\widehat{y}_{\widehat{T}_{-,+}} \geq \widehat{y}_+ \}\big)\big) \,+\, \widehat{y}'_+\mathbb{P}\big(\{\widehat{y}_{\widehat{T}_{-,+}} \geq \widehat{y}_+ \}\big)}{\mathbb{E}(\chi)}\,.
\end{equation}
Now an expression for $\mathbb{P}\big(\{ \widehat{y}_{\widehat{T}_{-,+}} \geq \widehat{y}_+ \}\big)$ is needed. A standard technique (see {\it e.g.}~\cite{Eth}) is based on the following lemma.

%%%%%%%%%%%%%%%%%%%%%%%%%%%%%%%%%%%%%%%%%
\begin{lemma}
\label{lemma:nu}
There is a unique solution $\nu \in (0,\infty)$ for the equation $\mathbb{E}(\kappa_0^{\nu}) = 1$, implying that $e^{C_0\nu\widehat{y}_n}$ is a martingale.
\end{lemma}
%%%%%%%%%%%%%%%%%%%%%%%%%%%%%%%%%%%%%%%%%

\noindent {\bf  Proof.} The main Hypothesis states that $\mathbb{P}(\{\chi > 0\}) > 0$, hence $\lim\limits_{\rho \to \infty} \mathbb{E}(e^{C_0\rho\chi}) = \infty$. Consider the map $\rho \in \mathbb{R} \mapsto \mathbb{E}(e^{C_0\rho\chi}) \in (0,\infty)$, which is differentiable at $\rho = 0$ with derivative
$$
\partial_{\rho}\;\mathbb{E}(e^{C_0\rho\chi})\big|_{\rho = 0} 
\;=\; 
\mathbb{E} (\log(\kappa_0)) 
\;<\; 0\,.
$$
This map is continuous, so the intermediate value theorem applies on $[0,\infty)$, yielding a solution $\nu$ for $\rho$ of $\mathbb{E}(e^{C_0\rho\chi}) = 1$ on $(0,\infty)$. As the map is strictly convex, this solution is unique.
\hfill $\square$

\vspace{.2cm}

As $\widehat{y}_n \in [\widehat{y}_--1, \widehat{y}_++1]$ and $|\chi| \leq 1$ a.s., the increments of the martingale $e^{C_0\nu\widehat{y}_n}$ of Lemma~\ref{lemma:nu} are uniformly bounded by $|e^{C_0\nu\widehat{y}_n} - e^{C_0\nu\widehat{y}_{n+1}}| = e^{C_0\nu\widehat{y}_n}|e^{C_0\nu\chi_n} - 1| \leq e^{C_0\nu(\widehat{y}_++1)}|e^{C_0\nu} - 1|$. Using $\mathbb{E}(\widehat{T}_{-,+}) < \infty$ from Lemma~\ref{lemma:E-T-slower} to apply the optional stopping theorem to this martingale yields
$$
1 
\;=\; 
e^{C_0\nu \cdot 0} 
\;=\; 
e^{C_0\nu \widehat{y}_0} 
\;=\; 
\mathbb{E}(e^{C_0\nu\widehat{y}_{\widehat{T}_{-,+}}}) 
\;=\; 
e^{C_0\nu \widehat{y}''_-}\big(1 - \mathbb{P}(\{ \widehat{y}_{\widehat{T}_{-,+}} \geq \widehat{y}_+ \})\big) + e^{C_0\nu \widehat{y}''_+}\mathbb{P}(\{ \widehat{y}_{\widehat{T}_{-,+}} \geq \widehat{y}_+ \})
\,.
$$
Inserting~\eqref{eq:slower-bis} into~\eqref{eq:slower} and combining this with the foregoing finally gives
\begin{align*}
\big(\mathbb{E}(\widehat{T}_1)\big)^{-1} 
&\;=\;
\left[\frac{\widehat{y}'_- + \mathbb{E}(\chi)}{\mathbb{E}(\chi)\mathbb{P}\big(\big\{\widehat{y}_{\widehat{T}_{-,+}} \geq \widehat{y}_+\big\}\big)}\, +\,\frac{\widehat{y}'_+ - \widehat{y}'_-}{\mathbb{E}(\chi)}\, +\, 2\right]^{-1}\\
&\;=\; 
\left[\left(1 + \frac{C_0\,\widehat{y}'_-}{\mathbb{E}(\log(\kappa_0))}\right)\,\frac{e^{C_0\nu\widehat{y}''_+}-e^{C_0\nu\widehat{y}''_-}}{1-e^{C_0\nu\widehat{y}''_-}}\, +\, \frac{C_0\,(\widehat{y}'_+ - \widehat{y}'_-)}{\mathbb{E}(\log(\kappa_0))}\, +\, 2\right]^{-1}
\,,
\end{align*}
which together with the two statements of Lemma~\ref{lemma:slower-y-lims} implies that
$$
C_- 
\;:=\;
\left(1 + \frac{C_0\,(\widehat{y}_--1)}{\mathbb{E}(\log(\kappa_0))}\right)^{-1}\frac{1-e^{C_0\nu(\widehat{y}_--1)}}{e^{C_0\nu}} 
$$
satisfies
$$
C_-
\;\leq\; 
\lim_{\epsilon \to 0} \left(1 + \frac{C_0\,\widehat{y}'_-}{\mathbb{E}(\log(\kappa_0))}\right)^{-1}\,\frac{1-e^{C_0\nu\widehat{y}''_-}}{\epsilon^{\nu}e^{C_0\nu\widehat{y}''_+}} = \lim_{\epsilon \to 0}  \frac{\big(\mathbb{E}(\widehat{T}_1)\big)^{-1}}{\epsilon^{\nu}}\,,
$$
namely the first statement of Proposition~\ref{prop:T}.

%%%%%%%%%%%%%%%%%%%%%%%%%%%%%%%%%%%%%%%%%%%%%%%%%%%%%%%%
\section{Upper bound on the rotation number}
\label{sec:upper}

This section is structured just as the previous one, namely first a faster comparison process satisfying~\eqref{stat:slower-faster} and~\eqref{stat:slower-faster-bis} is constructed and then the $\epsilon$-dependence of its expected stopping time is analyzed in order to prove the second statement of Proposition~\ref{prop:T}. It will be useful to introduce a suitable positive-valued function $\lambda$ of $\epsilon$, satisfying the defining properties
\begin{equation}
\label{eq:lambda}
\lim_{\epsilon \to 0}\; \lambda \;=\; 0\,,\qquad \lim_{\epsilon \to 0}\; \frac{\log(\lambda)}{\log(\epsilon)} \;=\; 0\,,
\qquad
\lim_{\epsilon \to 0} \;\frac{\epsilon}{\lambda} \;=\; 0
\,,	
\end{equation}
where the last property actually follows from the first two. For conciseness an additional notation is introduced:
$$
\Lambda \;:=\; e^{2C_0\lambda}\,.
$$
Note that $\Lambda$ also depends on $\epsilon$. Similar as in Section~\ref{sec:lower}, three reference points $0<\widetilde{x}_-<\widetilde{x}_c<\widetilde{x}_+<\infty$ will be needed w.r.t. which the dynamics has uniform properties schematically described in  Figure~\ref{fig:faster}. While similar to  Figure~\ref{fig:slower}, note that the original dynamics now is bounded above by these points, see Lemma~\ref{lemma:faster} below. For its proof, let us start out with a counterpart to Lemma~\ref{lemma:Q-lower}. 

%%%%%%%%%%%%%%%%%%%%%%%%%%%
\begin{lemma}
\label{lemma:Q-upper}
There exist $\widetilde{x}_-$ and $\widetilde{x}_+$ depending on $C_0$, $C_2$, $C_3$ and $\epsilon$ such that
$$
x \,\in \,[\widetilde{x}_-,\widetilde{x}_+]
\qquad
\Longrightarrow
\qquad
Q \cdot (D \cdot x) \,\leq\, \Lambda(D \cdot x)
\,,
$$ 
as well as
\begin{equation}
\label{eq:faster-x-lims}
\lim_{\epsilon \to 0} \;\frac{\lambda\,\widetilde{x}_-}{\epsilon} 
\;=\;
\frac{C_2\,e^{2C_0}}{2C_0}
\,,
\qquad
\lim_{\epsilon \to 0} \;\frac{\epsilon\,\widetilde{x}_+}{\lambda}
\;=\;
\frac{2C_0}{C_2\,e^{2C_0}}
\,.
\end{equation}
\end{lemma}
%%%%%%%%%%%%%%%%%%%%%%%%%%%

\noindent\textbf{Proof.}
For $x \in [0,+\infty)$ one can estimate
$$
Q \cdot (D \cdot x) 
\;=\; 
\frac{(1+\epsilon^2\alpha)(D \cdot x) + (a-b-\epsilon\beta)\epsilon}{1+\epsilon^2\delta - (a+b+\epsilon\gamma)\epsilon(D \cdot x)} 
\;\leq\; 
\frac{(1+C_3\epsilon^2)(D \cdot x) + (C_2+C_3\epsilon)\epsilon}{1-C_3\epsilon^2 - (C_2+C_3\epsilon)\epsilon(D \cdot x)}
\,.
$$
The latter is smaller than or equal to $\Lambda(D \cdot x)$ if and only if
$$
\Lambda(C_2+C_3\epsilon)\epsilon(D \cdot x)^2 - \big(\Lambda-1-C_3\epsilon^2(\Lambda+1)\big)(D \cdot x) + (C_2+C_3\epsilon)\epsilon 
\;\leq\; 
0\,.
$$
Let us equate this to $A(D \cdot x)^2 - B(D \cdot x) + C$, namely set
$$
A 
\;:=\; \Lambda(C_2+C_3\epsilon)\epsilon
\,,
\qquad 
B 
\;:= \;
\Lambda-1-C_3\epsilon^2(\Lambda+1)
\,,
\qquad 
C 
\;:= \;
(C_2+C_3\epsilon)\epsilon
\,.
$$
Note that for $\epsilon \to 0$, $\frac{A}{C_2\epsilon}$, $\frac{B}{2C_0\lambda}$ and $\frac{C}{C_2\epsilon}$ all converge to $1$ by~\eqref{eq:lambda}, hence $A,B,C \in (0,\infty)$ for $\epsilon$ small. Now let us search for real solutions of the quadratic equation $A(D \cdot x)^2 - B(D \cdot x) + C=0$. They clearly exist whenever $B^2 - 4AC \geq 0$, which follows from the foregoing limits and~\eqref{eq:lambda}. Then denote the two real zeroes of the quadratic equation by $x_- \leq x_+$, so $A(D \cdot x)^2 - B(D \cdot x) + C$ then equals $A\left[(D \cdot x)-x_+\right]\left[(D \cdot x)-x_-\right]$. The fact that $\sqrt{1-r} \geq 1-\frac{r}{2}-\frac{r^2}{2} \geq 1-r$ for all $r \in [0,1]$ implies after some algebra that $x_- \leq \frac{(B^2+4AC)C}{B^3}$ and $x_+\geq\frac{B^2-2AC}{AB} $. Hence let us set
\begin{align*}
\widetilde{x}_- 
&\;:=\; 
e^{2C_0}\,\frac{(B^2+4AC)C}{B^3}
\;=\;
\frac{\big[\big(\Lambda-1-C_3\epsilon^2(\Lambda+1)\big)^2 + 4\Lambda(C_2+C_3\epsilon)^2\epsilon^2\big](C_2+C_3\epsilon)\,e^{2C_0}\,\epsilon}{\left(\Lambda-1-C_3\epsilon^2\left[\Lambda+1\right]\right)^3}\,,\\
\widetilde{x}_+ 
&
\;:=\;
e^{-2C_0}\,\frac{B^2-2AC}{AB}
\;=\;
\frac{\big(\Lambda-1-C_3\epsilon^2(\Lambda+1)\big)^2 \,- \,2\,\Lambda\,(C_2+C_3\epsilon)^2\,\epsilon^2}{\big(\Lambda-1-C_3\epsilon^2(\Lambda+1)\big)\,\Lambda\,(C_2+C_3\epsilon)\,e^{2C_0}\,\epsilon}\,.
\end{align*}
For $x \in [\widetilde{x}_-,\widetilde{x}_+]$, one has due to $e^{-2C_0}x \leq (D \cdot x) \leq e^{2C_0}x$ that $(D \cdot x) \in [e^{-2C_0}\widetilde{x}_-,e^{2C_0}\widetilde{x}_+]\subset[x_-,x_+]$, which by the above implies the first statement. The limits~\eqref{eq:faster-x-lims} follow again by the given limit behavior of $A$, $B$ and $C$ as well as from~\eqref{eq:lambda}.
\hfill $\square$

\vspace{.2cm}

Let us now complete the left part of Figure~\ref{fig:faster} by setting
$$
\widetilde{x}_c
\;:=\;
e^{2C_0}\,\Lambda\,\widetilde{x}_-
\,.
$$
The next statement corresponds to Lemma~\ref{lemma:slower}.

\begin{figure}
	\begin{center}
		\begin{tikzpicture}[line join = round, line cap = round]
		\coordinate (a) at (0.0,0.3);
		\coordinate (b) at (0.0,-0.2);
		\coordinate (l) at (-0.2,0.0);
		\coordinate (l0) at (-6,0);
		\coordinate (l0) at (-6.5,0);
		\tick{(l0)};
		\coordinate (l0a) at ($(l0) + (a)$);
		\coordinate[label=below:{$0$}] (l0b) at ($(l0) + (b)$);
		\coordinate (l1) at (-5.5,0);
		\tick{(l1)};
		\coordinate (l1a) at ($(l1) + (a)$);
		\coordinate (l1al) at ($(l1) + (a) + (l)$);
		\draw[->] (l0a) to[out=90, in=75, looseness=1.5] node[midway,above,inner sep=4pt] {\eqref{stat:faster-start}} (l1al);
		\coordinate[label=below:{$\widetilde{x}_-$}] (l1b) at ($(l1) + (b)$);
		\coordinate (l2) at (-4.5,0);
		\tick{(l2)};
		\coordinate (l2al) at ($(l2) + (a) + (l)$);
		\draw[->] (l1a) to[out=90, in=75, looseness=1.5] node[midway,above,inner sep=4pt] {\eqref{stat:faster-between}} (l2al);
		\coordinate[label=below:{$\widetilde{x}_c$}] (l2b) at ($(l2) + (b)$);
		\coordinate (l3) at (-4,0);
		\coordinate (l4) at (-3,0);
		\coordinate (l5) at (-2.5,0);
		\tick{(l5)};
		\coordinate (l5a) at ($(l5) + (a)$);
		\coordinate[label=below:{$\widetilde{x}_+$}] (l5b) at ($(l5) + (b)$);
		\coordinate (lr) at (-1.5,0);
		\coordinate[label=left:{$\widetilde{x}$}] (lrr) at (-1,0);
		\coordinate (lrra) at ($(lrr) + (a)$);
		\draw[->] (l5a) to[out=90, in=120, looseness=1.5] node[midway,above,inner sep=4pt] {\eqref{stat:faster-end}} (lrra);
		\draw [-,color=black,line width=0.3mm] (l0)--(l3);
		\draw [-,color=black,dotted,line width=0.3mm] (l3)--(l4);
		\draw [->,color=black,line width=0.3mm] (l4) -- (lr);
		\coordinate (ml) at (-0.5,-0.5);
		\coordinate (mr) at (0.5,-0.5);
		\draw[->] (ml) to[bend right] node[midway,above,inner sep=4pt] {$\widetilde{f}$} (mr);
		\coordinate (rl) at (1.5,0);
		\coordinate (r1) at (2,0);
		\tick{(r1)};
		\coordinate[label=below:{$\widetilde{y}_-$}] (r1b) at ($(r1) + (b)$);
		\coordinate (r2) at (3,0);
		\tick{(r2)};
		\coordinate[label=below:{$0$}] (r2b) at ($(r2) + (b)$);
		\coordinate (r3) at (3.5,0);
		\coordinate (r4) at (4.5,0);
		\coordinate (r5) at (5,0);
		\tick{(r5)};
		\coordinate[label=below:{$\widetilde{y}_+$}] (r5b) at ($(r5) + (b)$);
		\coordinate[label=right:{$\widetilde{y}$}] (rr) at (5.5,0);
		\draw [-,color=black,line width=0.3mm] (rl)--(r3);
		\draw [-,color=black,dotted,line width=0.3mm] (r3)--(r4);
		\draw [->,color=black,line width=0.3mm] (r4) -- (rr);
		\end{tikzpicture}
		\caption{\it The arrows on the left part illustrate properties of the original Dyson-Schmidt dynamics on $(0,\infty)$ as stated in {\rm Lemma~\ref{lemma:faster}}. The right part illustrates the notations after the logarithmic transformation $\widetilde{f}$ to $\mathbb{R}$.}
		\label{fig:faster}
	\end{center}
\end{figure}

%%%%%%%%%%%%%%%%%%%%%%%%%%%%%%%%%%%%%%%
\begin{lemma}
\label{lemma:faster}
For each realization, one has
\begin{alignat}{3}
	& x \,\notin\, [0,\infty)  &\qquad \Longrightarrow \qquad & Q \cdot (D \cdot x) \,\notin\, [\widetilde{x}_-,\infty)\,,
	\label{stat:faster-start}\\
	& x \,\notin\, [\widetilde{x}_-,\infty)  &\qquad \Longrightarrow \qquad  &  Q \cdot (D \cdot x) \,\notin\, [\widetilde{x}_c,\infty)\,,
	\label{stat:faster-between}\\
	& Q \cdot (D \cdot x) \,\notin\, [0,\infty)  &\qquad \Longrightarrow \qquad  & x \,\notin\, [0,\widetilde{x}_+)\,.
	\label{stat:faster-end}
\end{alignat}
\end{lemma}
%%%%%%%%%%%%%%%%%%%%%%%%%%%%%%%%%%%%%%%

\noindent {\bf Proof.}
For~\eqref{stat:faster-start}, let $x \notin [0,\infty)$, so then $D \cdot x \notin [0,\infty)$. By combining the order-preserving property~\eqref{stat:order} with~\eqref{eq:Q-D-explicit} and the Hypothesis, one has for nonnegative $Q \cdot (D \cdot x)$ that
$$
Q \cdot (D \cdot x) 
\;<\;
Q \cdot 0 
\;=\; 
\tfrac{(a-b-\epsilon\beta)\epsilon}{1+\epsilon^2\delta} 
\;\leq\; 
\tfrac{(C_2+C_3\epsilon)\epsilon}{1-C_3\epsilon^2} 
\;\leq\; 
C_2\,e^{2C_0}\epsilon
\,.
$$
By~\eqref{eq:lambda} and~\eqref{eq:faster-x-lims} it indeed follows that $C_2e^{2C_0}\epsilon < \widetilde{x}_-$ for $\epsilon$ small enough. For the proof of~\eqref{stat:faster-between}, combining its hypothesis, the order-preserving property~\eqref{stat:order}, Lemma~\ref{lemma:Q-upper} and the fact that $D \cdot x' \leq e^{2C_0}x'$ for nonnegative $x'$, yields
$$
Q \cdot (D \cdot x) 
\;<\; 
Q \cdot (D \cdot \widetilde{x}_-)
\;\leq\; 
\Lambda \,e^{2C_0}\,\widetilde{x}_-
\;=\;\widetilde{x}_c
\,.
$$
Finally let us verify~\eqref{stat:faster-end} by contraposition. If $x \in [0,\widetilde{x}_+)$, then the order-preserving property~\eqref{stat:order} and Lemma~\ref{lemma:Q-upper} imply
$$
0 
\;\leq\; 
Q \cdot 0
\;\leq\;
Q \cdot (D \cdot x) 
\;\leq\; 
Q \cdot (D \cdot \widetilde{x}_+)
\;\leq\; 
\Lambda(D \cdot \widetilde{x}_+)
\;<\;\infty
\,,
$$ 
just as claimed.
\hfill $\square$

\vspace{.2cm}

Now a new process $\widetilde{x}=(\widetilde{x}_n)_{n\geq 0}$ on $[0,\infty]$ is constructed by setting for $n\geq 0$
$$
\widetilde{x}_0
\;=\;
\widetilde{x}_-\,,
\qquad
\widetilde{x}_{n+1} 
\;=\; 
\begin{cases}
\widetilde{x}_c\;, & \text{ if } \widetilde{x}_n \leq \widetilde{x}_-\,,
\\
\Lambda(D_n \cdot \widetilde{x}_n) \;,&\text{ if } \widetilde{x}_n \in (\widetilde{x}_-,\widetilde{x}_+)\;,
\\
\infty\;, &\text{ else, so if } \widetilde{x}_n \geq  \widetilde{x}_+\,.
\end{cases}
$$
Comparing with~\eqref{eq:Q-D-x}, the main case $\widetilde{x}_{n+1}=\Lambda(D_n \cdot \widetilde{x}_n)$ of this process merely bounds the action of $Q_n$ for $n\geq 2$. Let us now argue why this process satisfies the first inequality in~\eqref{stat:slower-faster}. Indeed, replacing the action of $Q_n$ by a multiplication by $\Lambda$ speeds up the process because of the order-preserving property~\eqref{stat:order} and Lemma~\ref{lemma:Q-upper} which applies in the case $\widetilde{x}_n \in (\widetilde{x}_-,\widetilde{x}_+)$. Carefully analyzing the first case in the definition of $\widetilde{x}_{n+1}$ in combination with~\eqref{stat:faster-start} and~\eqref{stat:faster-between} shows that a.s. $x_{N_{(1)} + n} \leq \widetilde{x}_n$ for all $n \in \lbrace 0,1,\dots,N_{(2)} - 1 - N_{(1)}\rbrace$, that is, as long as $x_{N_{(1)} + n} \in [0,\infty)$. Moreover, by~\eqref{stat:faster-end} it is indeed impossible that $\widetilde{x}_{N_{(2)} - N_{(1)}} \neq \infty$, as this would imply the contradiction $x_{N_{(2)}-1} \geq  \widetilde{x}_+ > \widetilde{x}_{N_{(2)} - N_{(1)}-1}$. Therefore also~\eqref{stat:slower-faster-bis} holds, and conversely indeed $\widetilde{T}_1 \leq N_{(2)} - N_{(1)}$ since a.s. $\widetilde{x}_{\widetilde{T}_1} = \infty$.

\vspace{.2cm}

Now let us proceed proving the limit behavior of $\big(\mathbb{E}(\widetilde{T}_1)\big)^{-1}$ as given in Proposition~\ref{prop:T}. As in the previous section, this is achieved by passing to a shifted logarithm of the Dyson-Schmidt variables, here via the map $\widetilde{f}:(0,\infty)\to \mathbb{R}$ given by 
$$
\widetilde{f}(x)
\;:=\;
\frac{1}{2C_0}\,\log\Big(\frac{x}{\widetilde{x}_c}\Big)
\,.
$$
By construction, $\widetilde{f}(\widetilde{x}_c) = 0$. Furthermore, for $n$ such that $\widetilde{x}_{n+1}<\infty$, let us introduce 
$$
\widetilde{y}_n
\;:=\;
\widetilde{f}(\widetilde{x}_{n+1})
\,,
\qquad
\widetilde{y}_-\;:=\;\widetilde{f}(\widetilde{x}_-)
\,,
\qquad
\widetilde{y}_+\;:=\;\widetilde{f}(\widetilde{x}_+)
\,,
$$
and the stopping time
$$
\widetilde{T}_{-,+}
\;:=\;
\inf\big\{ n \in \mathbb{N} \,:\, 
\widetilde{y}_n\notin (\widetilde{y}_-,\widetilde{y}_+) 
\big\}\,.
$$
As long as $n \leq \widetilde{T}_{-,+}$, it holds that
\begin{equation}
\label{eq:y-faster}
\widetilde{y}_n
\;=\; 
\tfrac{1}{2C_0}\log\Big(\frac{\Lambda^{n} (D^n \cdot \widetilde{x}_1)}{\widetilde{x}_c}\Big)
\;=\; 
\tfrac{1}{2C_0}\log\Big(\prod_{j=N_{(1)}+1}^{N_{(1)}+n} e^{2C_0\lambda}\kappa_j^2\Big) 
\;=\; 
\sum_{j=N_{(1)}+1}^{N_{(1)}+n} (\chi_j+\lambda)\,,
\end{equation}
namely $ \widetilde{y}_n $ is a random walk starting at $\widetilde{y}_0=0$. For $\lambda$ small enough, it  still contains a drift in the negative direction. The following two lemmata recollect properties about these newly introduced quantities.

%%%%%%%%%%%%%%%%%%%%%%%%%%%%%%%%%%%%%%%
\begin{lemma}
\label{lemma:faster-y-lims}
For $\epsilon \to 0$, both $\frac{C_0\widetilde{y}_+}{-\log(\epsilon)}$ and $-\widetilde{y}_-$ converge to $1$.
\end{lemma}
%%%%%%%%%%%%%%%%%%%%%%%%%%%%%%%%%%%%%%%

\noindent\textbf{Proof.}
The first statement follows from the limit behavior of $\widetilde{x}_-$ and $\widetilde{x}_+$ as given in~\eqref{eq:faster-x-lims}:
$$
\lim_{\epsilon \to 0} \frac{C_0\widetilde{y}_+}{-\log(\epsilon)} 
\;=\; 
\lim_{\epsilon \to 0} \frac{\log(\widetilde{x}_+)-\log (e^{-2C_0}\Lambda)-\log(\widetilde{x}_-)}{-2\log(\epsilon)} 
\;=\;
\lim_{\epsilon \to 0} \frac{\log(\lambda) - \log(\epsilon)}{-\log(\epsilon)}
\;=\;
1
$$
by~\eqref{eq:lambda}. The second statement follows from the observation that $\widetilde{y}_- = -1-\lambda$.
\hfill $\square$

\vspace{.2cm}

%%%%%%%%%%%%%%%%%%%%%%%%%%%%%%%%%%%%%%%
\begin{lemma}
\label{lemma:E-T-faster}
$\mathbb{E}(\widetilde{T}_{-,+}) < +\infty$.
\end{lemma}
%%%%%%%%%%%%%%%%%%%%%%%%%%%%%%%%%%%%%%%

\noindent\textbf{Proof.}
As $\lambda > 0$ and $\mathbb{P}(\{\chi > 0\})> 0$, it holds that $\widetilde{p}:=\mathbb{P}(\{\chi + \lambda > 0\})$ is strictly positive. Denoting $\widetilde{E} := \lceil\frac{\widetilde{y}_+-\widetilde{y}_-}{\lambda}\rceil$ and introducing the random variable
$$
\widetilde{N}
\;:=\;
\min\big\{  n \in \mathbb{N} \,:\, \chi_{(n-1)\widetilde{E} + 1} \geq \lambda\,,\;\; \chi_{(n-1)\widetilde{E} + 2} \geq \lambda\,, \,\dots\,,\;\; \chi_{n\widetilde{E}} \geq \lambda \big\}
\,,
$$ 
the latter then is geometrically distributed with success probability $\widetilde{p}^{\widetilde{E}}$. In particular, one has $\mathbb{E}(\widetilde{N}) < \infty$. Moreover, $\widetilde{T}_{-,+} < \widetilde{E}\,\widetilde{N}$ a.s. by construction, so $\mathbb{E}(\widetilde{T}_{-,+}) < \widetilde{E}\,\mathbb{E}(\widetilde{N}) < \infty$.
\hfill $\square$

\vspace{.2cm}

The connection between the two stopping times $\widetilde{T}_{-,+}$ and $\widetilde{T}_1$ is almost identical to that in the previous section: this time it holds that $\mathbb{E}\big(\widetilde{T}_1 \,\big|\, \widetilde{y}_{\widetilde{T}_{-,+}} \geq \widetilde{y}_+\big) = \mathbb{E}\big(\widetilde{T}_{-,+} + 2 \,\big|\, \widetilde{y}_{\widetilde{T}_{-,+}} \geq \widetilde{y}_+\big)$. Therefore, up to this single change, the argument leading to~\eqref{eq:slower} directly transposes (simply by replacing all hats with tildes, and with $2$ instead of $3$ everywhere), so that one has
\begin{equation}
\label{eq:faster}
\big(\mathbb{E}(\widetilde{T}_1)\big)^{-1}
\;=\;
\left[\frac{\mathbb{E}(\widetilde{T}_{-,+}) + 1}{\mathbb{P}\big(\big\{\widetilde{y}_{\widetilde{T}_{-,+}} \geq \widetilde{y}_+\big\}\big)}\, +\, 1\right]^{-1}\,.
\end{equation}
In complete analogy with the previous section, one can next define
\begin{align*}
\widetilde{y}'_-
&\;:=\;
\mathbb{E}\big(\widetilde{y}_{\widetilde{T}_{-,+}} \,\big|\, \widetilde{y}_{\widetilde{T}_{-,+}} \leq \widetilde{y}_-\big)\,,
&\qquad&
\widetilde{y}''_-
\;:=\;
\tfrac{1}{C_0\nu}\log\Big(\mathbb{E}\big(e^{C_0\nu\widetilde{y}_{\widetilde{T}_{-,+}}} \,\big|\, \widetilde{y}_{\widetilde{T}_{-,+}} \leq \widetilde{y}_-\big)\Big)\,,\\
\widetilde{y}'_+
&\;:=\;
\mathbb{E}\big(\widetilde{y}_{\widetilde{T}_{-,+}} \,\big|\, \widetilde{y}_{\widetilde{T}_{-,+}} \geq \widetilde{y}_+\big)\,,
&\qquad&
\widetilde{y}''_+
\;:=\;
\tfrac{1}{C_0\nu}\log\Big(\mathbb{E}\big(e^{C_0\nu\widetilde{y}_{\widetilde{T}_{-,+}}} \,\big|\, \widetilde{y}_{\widetilde{T}_{-,+}} \geq \widetilde{y}_+\big)\Big)\,,
\end{align*}
in which $\widetilde{y}'_-, \widetilde{y}''_- \in \left[\widetilde{y}_--1+\lambda,\widetilde{y}_-\right]$ and $\widetilde{y}'_+, \widetilde{y}''_+ \in \left[\widetilde{y}_+,\widetilde{y}_++1+\lambda\right]$. Now~\eqref{eq:y-faster} implies that this time $\widetilde{y}_n - n(\mathbb{E}(\chi) + \lambda)$ is a martingale. As $|\chi| \leq 1$ a.s., its increments are a.s. bounded, namely more precisely $|\widetilde{y}_{n+1} - (n+1)(\mathbb{E}(\chi) + \lambda) - \widetilde{y}_n + n(\mathbb{E}(\chi) + \lambda)| \leq 2$. With $\mathbb{E}(\widetilde{T}_{-,+}) < \infty$ from Lemma~\ref{lemma:E-T-faster}, the optional stopping theorem yields $0 = \mathbb{E}(\widetilde{y}_0 - 0 \cdot (\mathbb{E}(\chi) + \lambda)) = \mathbb{E}(\widetilde{y}_{\widetilde{T}_{-,+}} - \widetilde{T}_{-,+} \cdot (\mathbb{E}(\chi) + \lambda))$, or
\begin{equation}
\label{eq:faster-bis}
\mathbb{E}(\widetilde{T}_{-,+}) 
\;=\; 
\frac{\mathbb{E}\big(\widetilde{y}_{\widetilde{T}_{-,+}}\big)}{\mathbb{E}(\chi) + \lambda} 
\;=\; 
\frac{\widetilde{y}'_-\big(1 - \mathbb{P}\big(\{\widetilde{y}_{\widetilde{T}_{-,+}} \geq \widetilde{y}_+ \}\big)\big) \,+\, \widetilde{y}'_+\mathbb{P}\big(\{\widetilde{y}_{\widetilde{T}_{-,+}} \geq \widetilde{y}_+ \}\big)}{\mathbb{E}(\chi) + \lambda}\,.
\end{equation}
%

%%%%%%%%%%%%%%%%%%%%%%%%%%%%%%%%%%%%%%%%%
\begin{lemma}
\label{lemma:nu-tilde}
For $\lambda$ close enough to $0$, {\it i.e.} $\epsilon$ small enough, there is a unique solution $\widetilde{\nu} \in (0,\infty)$ for $\rho$ of the equation $\mathbb{E}(e^{C_0\rho(\chi + \lambda)}) = 1$, implying that $e^{C_0\widetilde{\nu}\widetilde{y}_n}$ is a martingale. Moreover, $\widetilde{\nu} < \nu$ and $\lim_{\lambda \to 0} \widetilde{\nu} = \nu$.
\end{lemma}
%%%%%%%%%%%%%%%%%%%%%%%%%%%%%%%%%%%%%%%%%

\noindent\textbf{Proof.} As $\mathbb{E}(\chi) + \lambda < 0$ for $\lambda$ small enough and $\mathbb{P}(\{\chi + \lambda > 0\})$ is still positive, the proof of existence and uniqueness of $\widetilde{\nu}$ is identical to that of Lemma~\ref{lemma:nu}. Now, for $\rho \in (0,\infty)$ the value of $e^{C_0\rho\lambda}$ strictly decreases as $\lambda \to 0$. The strict convexity thus implies that $\widetilde{\nu}$ is strictly increasing as $\lambda \to 0$, hence $\widetilde{\nu} < \nu$. If $\widetilde{\nu} \leq \nu^{\prime}$ for all $\lambda > 0$ and some $\nu^{\prime} < \nu$, then $\widetilde{\nu} < \frac{\nu^{\prime}+\nu}{2} < \nu$ so that $\mathbb{E}\big(\exp(C_0\frac{\nu^{\prime}+\nu}{2}(\chi + \lambda))\big) \geq 1$ for $\lambda$ sufficiently small, contradicting the uniqueness of the solution $\rho = \nu$ on $(0,\infty)$ of $\mathbb{E}(e^{C_0\rho\chi}) = 1$.
\hfill $\square$

\vspace{.2cm}

As $\widetilde{y}_n \in [\widetilde{y}_--1+\lambda, \widetilde{y}_++1+\lambda]$ and $|\chi| \leq 1$ a.s., the increments of the martingale $e^{C_0\widetilde{\nu}\widetilde{y}_n}$ of Lemma~\ref{lemma:nu-tilde} are uniformly bounded by $|e^{C_0\widetilde{\nu}\widetilde{y}_n} - e^{C_0\widetilde{\nu}\widetilde{y}_{n+1}}| = e^{C_0\widetilde{\nu}\widetilde{y}_n}|e^{C_0\widetilde{\nu}(\chi_n + \lambda)} - 1| \leq e^{C_0\widetilde{\nu}(\widetilde{y}_++1+\lambda)}|e^{C_0\widetilde{\nu}(1+\lambda)} - 1|$. Then, exactly as in the previous section, replacing $\chi$ by $\chi + \lambda$ and all hats by tildes, one gets
\begin{equation}
\label{eq:faster-ter}
\big(\mathbb{E}(\widetilde{T}_1)\big)^{-1}
\;=\;
\left[\left(1 + \frac{C_0\,\widetilde{y}'_-}{\mathbb{E}(\log(\kappa_0)) + C_0\lambda}\right)\,\frac{e^{C_0\widetilde{\nu}\widetilde{y}''_+}-e^{C_0\widetilde{\nu}\widetilde{y}''_-}}{1-e^{C_0\widetilde{\nu}\widetilde{y}''_-}}\, +\, \frac{C_0\,(\widetilde{y}'_+ - \widetilde{y}'_-)}{\mathbb{E}(\log(\kappa_0)) + C_0\lambda}\, +\, 1\right]^{-1}
\,,
\end{equation}
which together with the two statements of Lemma~\ref{lemma:faster-y-lims} implies the second statement of Proposition~\ref{prop:T} with
$$
C_+ \;:=\; \left(1 + \frac{C_0\,\widetilde{y}_-}{\mathbb{E}(\log(\kappa_0))}\right)^{-1}(1-e^{C_0\nu\widetilde{y}_-}) 
\;,
$$
because
$$
C_+
\;\geq \;
\lim_{\epsilon \to 0} \left(1 + \frac{C_0\,\widetilde{y}'_-}{\mathbb{E}(\log(\kappa_0)) + C_0\lambda}\right)^{-1}\,\frac{1-e^{C_0\widetilde{\nu}\widetilde{y}''_-}}{\epsilon^{\widetilde{\nu}}e^{C_0\widetilde{\nu}\widetilde{y}''_+}} 
\;=\;
\lim_{\epsilon \to 0}  \frac{\big(\mathbb{E}(\widetilde{T}_1)\big)^{-1}}{\epsilon^{\widetilde{\nu}}}\,.
$$
%

%%%%%%%%%%%%%%%%%%%%%%%%%%%%%%%%%%
\section{Modifications for the balanced case}
\label{sec:balanced}

\begin{figure}
	\begin{center}
		\begin{tikzpicture}[line join = round, line cap = round]
		\coordinate (a) at (0.0,0.8);
		\coordinate (b) at (0.0,-0.4);
		\coordinate (l) at (-1,0);
		\coordinate (m0) at (0,0);
		\tick{(m0)};
		\coordinate[label=above:{$\epsilon$}, label=below:{$\curvearrowright$}] (m0a) at ($(m0) + (a)$);
		\coordinate[label=below:{$0$}] (m0b) at ($(m0) + (b)$);
		\coordinate (m1) at (1.5,0);
		\tick{(m1)};
		\coordinate[label=above:{$\epsilon$}, label=below:{$\curvearrowright$}] (m1a) at ($(m1) + (a)$);
		\coordinate[label=below:{$\frac{\pi}{2}$}] (m1b) at ($(m1) + (b)$);
		\coordinate (m2) at (3,0);
		\tick{(m2)};
		\coordinate[label=above:{$\epsilon$}, label=below:{$\curvearrowright$}] (m2a) at ($(m2) + (a)$);
		\coordinate[label=below:{$\pi$}] (m2b) at ($(m2) + (b)$);
		\coordinate (m3) at (4.5,0);
		\tick{(m3)};
		\coordinate[label=above:{$\epsilon$}, label=below:{$\curvearrowright$}] (m3a) at ($(m3) + (a)$);
		\coordinate[label=below:{$\frac{3\pi}{2}$}] (m3b) at ($(m3) + (b)$);
		\coordinate (m4) at (6.0,0);
		\tick{(m4)};
		\coordinate[label=above:{$\epsilon$}, label=below:{$\curvearrowright$}] (m4a) at ($(m4) + (a)$);
		\coordinate[label=below:{$2\pi$}] (m4b) at ($(m4) + (b)$);
		\coordinate (m5) at (7.5,0);
		\tick{(m5)};
		\coordinate[label=above:{$\epsilon$}, label=below:{$\curvearrowright$}] (m5a) at ($(m5) + (a)$);
		\coordinate[label=below:{$\frac{5\pi}{2}$}] (m5b) at ($(m5) + (b)$);
		\coordinate[label=right:{$\theta$}] (r) at (8.5,0);
		\draw [->,color=black,line width=0.3mm] (l)--(r);
		\end{tikzpicture}
		\caption{\it The dynamics of $\theta_n$ on the real line in the balanced case where $\mathbb{E}(\log(\kappa_0)) = 0$. Contrary to the unbalanced case depicted in Figure~\ref{fig:theta-unbalanced}, there are no drifts in this situation.}
		\label{fig:theta-balanced}
	\end{center}
\end{figure}
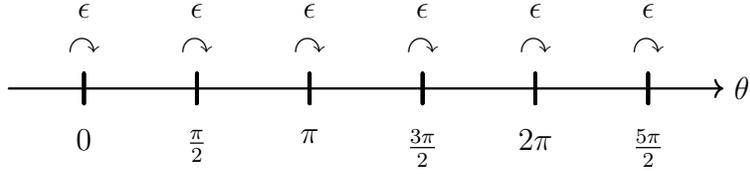

This final section considers in the balanced case $\mathbb{E}(\log(\kappa_0)) = 0$. Hence Figure~\ref{fig:theta-unbalanced} is not valid any longer, but rather has to be modified to Figure~\ref{fig:theta-balanced}. The action induced by $D_{\kappa}$ now yields no average drift everywhere on $\overline{\mathbb{R}}$. Therefore the random dynamics on the two half-axis $(-\infty,0)\cup\{\infty\}$ and $[0,\infty)$ is essentially the same, up to flipping the sign of $b_n$, swapping $\alpha^\epsilon_n$ for $\delta^\epsilon_n$ and $\beta^\epsilon_n$ for $-\gamma^\epsilon_n$, as well as changing $\kappa_n$ to $\kappa_n^{-1}$. Indeed, the bijective orientation preserving map $x \in(-\infty,0)\cup\{\infty\}\mapsto -x^{-1}\in[0,\infty)$ identifies these intervals and, moreover,
$$
Q^\epsilon_n \cdot (-x^{-1}) 
\;=\;
-\left[\frac{(1+\epsilon^2\delta^\epsilon_n)x + (a_n+b_n+\epsilon\gamma^\epsilon_n)\epsilon}{1+\epsilon^2\alpha^\epsilon_n - (a_n-b_n-\epsilon\beta^\epsilon_n)\epsilon x}\right]^{-1}\,, 
\quad 
D_n \cdot (-x^{-1}) 
\;=\; 
-\,\left[\kappa_n^{-2}x\right]^{-1}\,.
$$
As all estimates making use of the constants $C_1$, $C_2$, $C_3$ and $\mathbb{E}(\log(\kappa_0)^2)$ are invariant under the above swapping, it is sufficient to analyze the random dynamics on $[0,\infty)$. So,~\eqref{ineq:slower-faster} changes to
\begin{equation}
\label{ineq:slower-faster-balanced}
\frac{1}{2\mathbb{E}(\widehat{T}_1^\epsilon)} 
\;\leq\; 
\lim_{N \to \infty} \frac{1}{N}\frac{\mathbb{E}(\theta^\epsilon_N)}{\pi} 
\;\leq\; 
\frac{1}{2\mathbb{E}(\widetilde{T}_1^\epsilon)}\,,
\end{equation}
for which again families of nonnegative i.i.d. random variables $(\widehat{x}^\epsilon_{2k-1})_{k\in\mathbb{N}}$, $(\widehat{x}^\epsilon_{2k})_{k\in\mathbb{N}}$, $(\widetilde{x}^\epsilon_{2k-1})_{k\in\mathbb{N}}$ and $(\widetilde{x}^\epsilon_{2k})_{k\in\mathbb{N}}$ can be constructed. Note that this time all bounds do not differentiate between the processes with odd and even index $k$. Applying logarithmic transformations similar to $\widehat{f}$ and $\widetilde{f}$ to the processes $\widehat{x}^\epsilon_1$ and $\widetilde{x}^\epsilon_1$, one can obtain exactly the same processes $\widehat{y}$ and $\widetilde{y}$ as in~\eqref{eq:y-slower} and~\eqref{eq:y-faster} (again for $n$ smaller than some stopping time similar to $\widehat{T}_{-,+}$ or $\widetilde{T}_{-,+}$ respectively). Also the calculations leading to~\eqref{eq:slower} and~\eqref{eq:faster} remain valid. As in the previous two sections, each expectation in~\eqref{ineq:slower-faster-balanced} can be found by applying the optional stopping theorem to two martingales. In addition to the primed constants that were introduced before, let us set
\begin{align*}
\widehat{y}'''_-
&\;:=\;
-\sqrt{\mathbb{E}\big(\widehat{y}^2_{\widehat{T}_{-,+}} \,\big|\, \widehat{y}_{\widehat{T}_{-,+}} \leq \widehat{y}_-\big)}\,,
&\qquad&
\widetilde{y}'''_-
\;:=\;
-\sqrt{\mathbb{E}\big(\widetilde{y}^2_{\widetilde{T}_{-,+}} \,\big|\, \widetilde{y}_{\widetilde{T}_{-,+}} \leq \widetilde{y}_-\big)}\,,\\
\widehat{y}'''_+
&\;:=\;
\sqrt{\mathbb{E}\big(\widehat{y}^2_{\widehat{T}_{-,+}} \,\big|\, \widehat{y}_{\widehat{T}_{-,+}} \geq \widehat{y}_+\big)}\,,
&\qquad&
\widetilde{y}'''_+
\;:=\;
\sqrt{\mathbb{E}\big(\widetilde{y}^2_{\widetilde{T}_{-,+}} \,\big|\, \widetilde{y}_{\widetilde{T}_{-,+}} \geq \widetilde{y}_+\big)}\,.
\end{align*}
The estimates $\widehat{y}'''_- \in \left[\widehat{y}_--1,\widehat{y}_-\right]$, $\widetilde{y}'''_- \in \left[\widetilde{y}_--1+\lambda,\widetilde{y}_-\right]$, $\widehat{y}'''_+ \in \left[\widehat{y}_+,\widehat{y}_++1\right]$, $\widetilde{y}'''_+ \in \left[\widetilde{y}_+,\widetilde{y}_++1+\lambda\right]$ will again be used in what follows to bound these quantities. In the slower case (with hats), applying the optional stopping theorem to the martingales $\widehat{y}_n$ and $\widehat{y}_n^2 - n\mathbb{E}(\chi^2)$ then yields
$$
\big(\mathbb{E}(\widehat{T}_1)\big)^{-1} 
\;=\; 
\frac{\mathbb{E}(\chi^2)}{(\widehat{y}'''_+)^2}\left[1 \,+\, \frac{\mathbb{E}(\chi^2)(\widehat{y}'_+ - 3\widehat{y}'_-) + \widehat{y}'_+(\widehat{y}'''_-)^2}{-\widehat{y}'_-(\widehat{y}'''_+)^2}\right]^{-1}
\,,
$$
which together with the limits of Lemma~\ref{lemma:slower-y-lims} then shows the first statement of Proposition~\ref{prop:T-balanced}. In addition to the defining properties of $\lambda$ in~\eqref{eq:lambda}, it will be necessary to require
$$
\lim_{\epsilon \to 0} \lambda\log(\epsilon)
\;=\;
0
$$
to control the faster process (with tildes) in the balanced case, as this implies $\lim_{\epsilon \to 0} \lambda\widetilde{y}_+ = 0$. One can check that the choice $\lambda := [\log(\epsilon)]^{-2}$ meets all the given conditions. A first martingale is now given by $\widetilde{y}_n - n\lambda$ which leads to exactly the same result as~\eqref{eq:faster-bis} with $\mathbb{E}(\chi) = 0$ in this case. Next, similar to the proofs of Lemmata~\ref{lemma:nu} and~\ref{lemma:nu-tilde}, one can show that there exists a unique real solution $\widetilde{\rho}$ for $\rho$ solving $\mathbb{E}(e^{C_0\rho(\chi + \lambda)}) = 1$. This quantity must be negative, clearly depends on $\lambda$ (so on $\epsilon$) and obeys $\lim_{\lambda \to 0} \widetilde{\rho} = 0$. The implicit definition of $\widetilde{\rho}$ as a function of $\lambda$ can be written as $\lambda = \frac{\log(\mathbb{E}(e^{C_0\widetilde{\rho}\chi}))}{-C_0\widetilde{\rho}}$. By Fubini's theorem, this is an analytic function in $\widetilde{\rho}$ (as it is also well-defined for $\widetilde{\rho}$ positive, so $\lambda$ negative), with $\frac{-C_0\mathbb{E}(\chi^2)}{2} \neq 0$ as its first derivative w.r.t. $\widetilde{\rho}$. This allows to use the Lagrange inversion theorem for analytic functions, which shows
\begin{equation}
\label{eq:rho-tilde}
\widetilde{\rho}
\;=\;
-\frac{2}{C_0\,\mathbb{E}(\chi^2)}\,\lambda \,-\, \frac{4\,\mathbb{E}(\chi^3)}{3\,C_0\left(\mathbb{E}(\chi^2)\right)^3}\,\lambda^2 \,+\, \mathcal{O}(\lambda^3)\,.
\end{equation}
Inserting this after applying the optional stopping theorem to the martingale $e^{C_0\widetilde{\rho}\widetilde{y}_n}$ then yields
\begin{align*}
1
&\;=\;
\mathbb{E}\big(e^{C_0\widetilde{\rho} \cdot 0}\big)
\;=\;
\mathbb{E}\big(e^{C_0\widetilde{\rho}\widetilde{y}_0}\big)
\;=\;
\mathbb{E}\big(e^{C_0\widetilde{\rho}\widetilde{y}_{\widetilde{T}_{-,+}}}\big)\\
&\;=\;
1 \,-\, \mathbb{E}(\widetilde{y}_{\widetilde{T}_{-,+}})
\;\left[\frac{2\,\lambda}{\mathbb{E}(\chi^2)} + \frac{4\,\mathbb{E}(\chi^3)\,\lambda^2}{3\big(\mathbb{E}(\chi^2)\big)^2}\right]
\,+\,\frac{1}{2}\;\mathbb{E}(\widetilde{y}^2_{\widetilde{T}_{-,+}})\;\frac{4\,\lambda^2}{\big(\mathbb{E}(\chi^2)\big)^2}\\
&\qquad+ \;\mathcal{O}\left(\lambda^3\right) \cdot \big(1 - \mathbb{P}\big(\{ \widetilde{y}_{\widetilde{T}_{-,+}} \geq \widetilde{y}_+ \}\big)\big) 
\,+\, \mathcal{O}\big((\lambda\widetilde{y}_+)^3\big) \cdot \mathbb{P}\big(\{ \widetilde{y}_{\widetilde{T}_{-,+}} \geq \widetilde{y}_+ \}\big)\,,
\end{align*}
where the big $\mathcal{O}$-notation makes sense as it was required earlier that $\lambda\widetilde{y}_+ \to 0$ for $\epsilon \to 0$. Hence,
$$
\frac{1}{\mathbb{P}(\{ \widetilde{y}_{\widetilde{T}_{-,+}} \geq \widetilde{y}_+ \})}
\;=\; 
\frac{\widetilde{y}'_+-\widetilde{y}'_-}{-\widetilde{y}'_-}\left[1 - \frac{\big((\widetilde{y}'''_-)^2\widetilde{y}'_+ - \widetilde{y}'_-(\widetilde{y}'''_+)^2\big)\lambda}{-\widetilde{y}'_-(\widetilde{y}'_+-\widetilde{y}'_-)\mathbb{E}(\chi^2)} 
\,+\,\mathcal{O}\left([\lambda\widetilde{y}_+]^2\right)\right]\,,
$$
when carefully treating the error terms. Inserting this and~\eqref{eq:faster-bis} (with $\mathbb{E}(\chi) = 0$) into~\eqref{eq:faster} yields
\begin{align*}
\left(\mathbb{E}(\widetilde{T}_1)\right)^{-1} 
&\;=\;
\left[\frac{1}{\mathbb{P}(\{\widetilde{y}_{\widetilde{T}_{-,+}} \geq \widetilde{y}_+\})}\left[\frac{\widetilde{y}'_-}{\lambda} \,+\, 1\right]\, +\, \frac{\widetilde{y}'_+ - \widetilde{y}'_-}{\lambda}\, +\, 1\right]^{-1}\\
&\;=\;
\frac{\mathbb{E}(\chi^2)}{(\widetilde{y}''_+)^2}\left[1\; +\; \frac{(\widetilde{y}''_-)^2\widetilde{y}'_+\, +\, \mathbb{E}(\chi^2)(\widetilde{y}'_+ - 2\,\widetilde{y}'_-)}{-\widetilde{y}'_-(\widetilde{y}''_+)^2} + \mathcal{O}(\lambda\widetilde{y}_+)\right]^{-1}\,,
\end{align*}
which together with the limits stated in Lemma~\ref{lemma:faster-y-lims} implies the second statement of Proposition~\ref{prop:T-balanced}.

\vspace{.2cm}

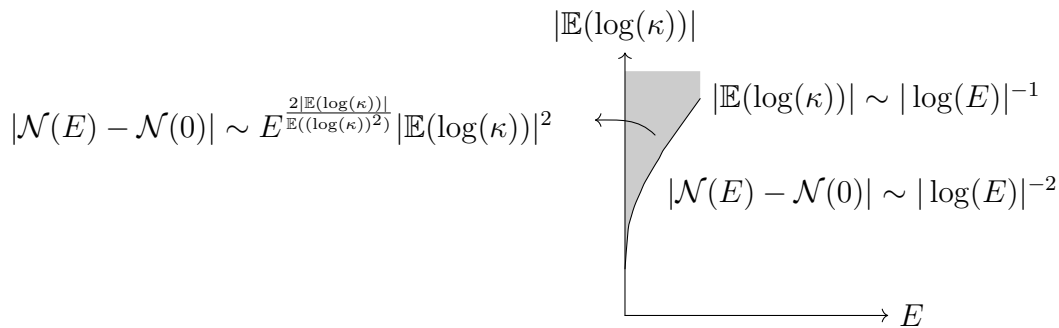
\begin{figure}
	\begin{center}
		\begin{tikzpicture}[line join = round, line cap = round]
		\draw[fill=gray!40,draw=none] plot[domain=0.0005:1.0,smooth] (\x,{-6.0/ln(\x/8.0)}) -- (1.0,3.25) -- (0,3.25) -- cycle node[midway,right] {$\quad |\mathcal{N}(E)-\mathcal{N}(0)| \sim |\log(E)|^{-2}$};
		\draw[domain=0.0005:1.0,smooth] plot (\x,{-6.0/ln(\x/8.0)}) node[right] {$|\mathbb{E}(\log(\kappa))| \sim |\log(E)|^{-1}$};
		\draw[->] (0.4,2.4) to[out=135, in=0] (-0.4,2.6) node[left] {$|\mathcal{N}(E)-\mathcal{N}(0)| \sim E^{\frac{2|\mathbb{E}(\log(\kappa))|}{\mathbb{E}((\log(\kappa))^2)}}|\mathbb{E}(\log(\kappa))|^2 \quad$};
		\draw[->] (0,0) -- (3.5,0) node[right] {$E$};
		\draw[->] (0,0) -- (0,3.5) node[above] {$|\mathbb{E}(\log(\kappa))|$};
		\end{tikzpicture}
		\caption{\it If $E$ and $|\mathbb{E}(\log(\kappa))|$ simultaneously tend to $0$, two types of scaling behavior for $|\mathcal{N}(E)-\mathcal{N}(0)|$ are separated by the curves on which $|\mathbb{E}(\log(\kappa))|$ is proportional to $|\log(E)|^{-1}$.}
		\label{fig:scaling-diagram}
	\end{center}
\end{figure}

\noindent {\bf Remark:} It is possible to analyze the scaling of the quantity $|\mathcal{N}(E)-\mathcal{N}(0)|$ when both $E$ and $|\mathbb{E}(\log(\kappa))|$ (or, equivalently, $|\mathbb{E}(\chi)|$) tend to zero. There are two different regimes, as depicted in Figure~\ref{fig:scaling-diagram}.  If $\lim_{E \to 0} |\mathbb{E}(\chi)\log(E)| < \infty$, then $|\mathcal{N}(E)-\mathcal{N}(0)|$ is proportional to $|\log(E)|^{-2}$. If the given limit equals zero, then the analysis in this section applies with $|\mathbb{E}(\chi)|$ taking the role of $\lambda$ (and $E$ that of $\epsilon$). For a non-vanishing limit, a lower (with hats instead of tildes and $\lambda$ set to zero) and an upper bound on $|\mathcal{N}(E)-\mathcal{N}(0)|$ are given by~\eqref{eq:faster-ter}, in which $\widetilde{\nu}$ needs to be replaced by $\widetilde{\rho}$. In its expansion~\eqref{eq:rho-tilde}, one then needs to replace $\lambda$ by $|\mathbb{E}(\chi)|$ and $|\mathbb{E}(\chi)|+\lambda$ respectively. If $\mathbb{E}(\chi)\log(E)$ converges to a non-zero constant (the case on the separating line in Figure~\ref{fig:scaling-diagram}), then also the factor $e^{C_0\widetilde{\rho}\widetilde{y}''_+}$ tends to a positive constant, and a further expansion in $|\mathbb{E}(\chi)|$ shows that $|\mathcal{N}(E)-\mathcal{N}(0)|$ is also in this case proportional to $|\log(E)|^{-2}$. Finally, if $|\mathbb{E}(\chi)\log(E)| \to \infty$ for $E \to 0$, then also $e^{C_0\widetilde{\rho}\widetilde{y}''_+} \to \infty$. The dominant term between brackets in~\eqref{eq:faster-ter} contains the latter factor, which then implies the scaling of $|\mathcal{N}(E)-\mathcal{N}(0)|\sim E^{\frac{2|\mathbb{E}(\log(\kappa))|}{\mathbb{E}((\log(\kappa))^2)}}|\mathbb{E}(\log(\kappa))|^2$ as indicated for the grey region in Figure~\ref{fig:scaling-diagram}. 
\hfill $\diamond$

\vspace{.2cm}

\noindent {\bf Acknowledgements:} The authors thank Sasha Sodin for bringing the works of Dyson \cite{Dys} and Kotowski and Vir\'ag \cite{KV} to their attention after a first preprint had appeared. This work was supported by the DFG grant SCHU 1358/8-1 and the Chilean grant FONDECYT 1201836. This manuscript has no associated data. The authors have no competing interests to declare that are relevant to the content of this article.

%%%%%%%%%%%%%%%%%%%%%%%%%%%%%%%%%%%%%%%%%%%%%

\end{document}